\documentclass{aa} 

\usepackage{graphicx}
\usepackage{mathabx} 
\usepackage{natbib}
\usepackage{txfonts}

\usepackage[citecolor=blue, urlcolor=blue, linkcolor=blue, colorlinks=true]{hyperref}

\usepackage[normalem]{ulem}

\begin{document}

   \title{Searching for H$_{\alpha}$-emitting sources in the gaps of five transitional disks\thanks{Based on observations obtained at Paranal Observatory under programs 100.C-0182(A) and 101.C-0078(A).}}
   \subtitle{SPHERE/ZIMPOL high-contrast imaging}
   
   \author{N. Hu\'elamo\inst{1}
          \and
          G. Chauvin\inst{2,3}
          \and
          I. Mendigut\'{\i}a\inst{1}
          \and 
          E. Whelan\inst{4}
          \and
          J. M. Alcal\'a\inst{5}
          \and 
          G. Cugno\inst{6}
          \and
          H. M. Schmid\inst{6} 
          \and
          I. de Gregorio-Monsalvo\inst{7}
          \and
          A. Zurlo\inst{8,9}
          \and
          D. Barrado\inst{1}
          \and
           M. Benisty\inst{2,3}
           \and
           S. P. Quanz\inst{6}
           \and 
          H. Bouy\inst{10}
          \and
          B. Montesinos\inst{1}
 \and
  Y. Beletsky\inst{11}
  \and
          J. Szulagyi\inst{6}
           }  
          \institute{Centro de Astrobiolog\'{\i}a (CAB), CSIC-INTA, ESAC Campus, Camino bajo del Castillo s/n, E-28692 Villanueva de la Ca\~nada, Madrid, Spain\\
              \email{nhuelamo@cab.inta-csic.es}
             \and
             Unidad Mixta Internacional Franco-Chilena de Astronom\'{i}a, CNRS/INSU UMI 3386 and Departamento de Astronom\'{i}a, Universidad de Chile, Casilla 36-D, Santiago, Chile
             \and
             Univ. Grenoble Alpes, CNRS, IPAG, F-38000 Grenoble, France
             \and
             Maynooth University Department of Experimental Physics, National University of Ireland, Maynooth Co. Kildare, Ireland
             \and
             INAF-Osservatorio Astronomico di Capodimonte, via Moiariello 16, 80131 Napoli, Italy
             \and
             ETH Zurich, Institute of Particle Physics and Astrophysics, Wolfgang-Pauli-Strasse 27, 8093 Zurich, Switzerland
             \and
             European Southern Observatory, Alonso de Cordova 3107, Casilla 19, Vitacura, Santiago, Chile
               \and
 N\'ucleo de Astronom\'{\i}a, Facultad de Ingenier\'{\i}a y Ciencias, Universidad Diego Portales, Av. Ejercito 441, Santiago, Chile
  \and
  Escuela de Ingenier\'{\i}a Industrial, Facultad de Ingenier\'ia y Ciencias, Universidad Diego Portales, Av. Ejercito 441, Santiago, Chile
  \and
             Laboratoire d'Astrophysique de Bordeaux, Univ. Bordeaux, CNRS, B18N, all\'ee Geoffroy Saint-Hilaire, 33615 Pessac, France
\and
             Las Campanas Observatory, Carnegie Institution of Washington, Colina el Pino, Casilla 601 La Serena, Chile
             }   
           \date{Received ; accepted }

 \abstract
   {(Pre-)Transitional disks show gaps and cavities that can be related to ongoing planet formation. According to theory, young embedded planets can accrete material from the circumplanetary and circumstellar disks and can be detected using accretion tracers, such as the H$_{\alpha}$ emission line.}
   {We aim to detect accreting protoplanets within the cavities of five (pre-)transitional disks through adaptive-optics(AO)-assisted spectral angular differential imaging in the optical regime.}
   {We performed simultaneous AO observations in the H$_{\alpha}$ line and the adjacent continuum using the Spectro-Polarimetric High-contrast Exoplanet REsearch (SPHERE) with the Zurich Imaging Polarimeter (ZIMPOL) at the Very Large Telescope (VLT). We combined spectral and angular differential imaging techniques to increase the contrast in the innermost regions close to the star and search for the signature of young accreting protoplanets.}
   {The reduced images show no clear H$_{\alpha}$ point source around any of the targets. We report the presence of faint H$_{\alpha}$ emission around TW~Hya and HD163296: while the former is most probably an artifact related to a spike, the nature of the latter remains unclear. The spectral and angular differential images yield contrasts of 6--8\,magnitudes at $\sim$100\,mas from the central stars, except in the case of LkCa15, with values of $\sim$3\,mag. We used the contrast curves to estimate average upper limits to the H$_{\alpha}$ line luminosity of $L_{\rm H_{\alpha}}\sim$ 5$\times$10$^{-6}$ $L_{\odot}$ at separations $\ge$200\,mas for TW\,Hya, RXJ1615, and T Cha, while for HD163296 and LkCa15 we derive values of $\sim 3\times$10$^{-5}$ $L_{\odot}$.
   We estimated upper limits to the accretion luminosity of potential protoplanets, obtaining that planetary models provide an average value of $L_{\rm acc} \sim 10^{-4}$ $L_{\odot}$ at 200\,mas, which is about two orders of magnitude higher than the $L_{\rm acc}$ estimated from the extrapolation of the  $L_{H_{\alpha}}$ -- $L_{acc}$ stellar relationship.} 
   {When considering all the objects observed with SPHERE/ZIMPOL in the H$_{\alpha}$ line, 5 in this work and 13 from the literature, 
   we can explain the lack of protoplanet detections by a  combination of factors, such as a majority of low-mass, low-accreting planets; potential episodic accretion; significant extinction from the circumstellar and circumplanetary disks; and the fact that the contrast is less favorable at separations of smaller than 100\,mas, where giant planets are more likely to form.}
  \keywords{stars: pre-main sequence  --  stars: planetary systems -- planets and satellites: detection -- planet--disk interactions -- stars: individual: RXJ1615, TW~Hya, T~Cha, LkCa15, HD163296 -- techniques: high angular resolution}
\maketitle

\section{Introduction}

Circumstellar disks are the cradle of planetary systems.
High-angular-resolution observations of these disks have revealed a plethora of structures, including spiral arms, warps, gaps, and radial streams 
that might be related to ongoing planetary formation \citep[e.g.][]{Andrews2011,Mayama2012,Grady2013,Boccaletti2013,Pinilla2015,Benisty2015,Perez2016}. 
 Great efforts have been made to detect protoplanets embedded in these disks using different observational techniques. As a result, there are now two confirmed protoplanets around the star PDS\,70 \citep{Keppler2018,Mesa2019}, one around AB~Aur \citep{Currie2022}, and a significant number of candidates \citep[e.g.,][]{Biller2014,Reggiani2018,Tsu2019,Bocca2020,Pinte2020}. 

Planet-formation theories predict that embedded giant planets are surrounded by  circumplanetary disks (CPDs) from which they accrete material \citep[e.g.,][]{Lovelace2011,Gressel2013,Szula2014}. 
The accretion of matter can take place via different mechanisms. For example, in the presence of a magnetic field, magnetospheric accretion could work as in the case of low-mass stars \citep{Lovelace2011,Zhu2015}. In this scenario, the material is expected to reach the planet surface following the magnetic field lines. Another mechanism is ``boundary layer accretion'' \citep[e.g.,][]{Owen2016}, whereby material is accreted directly onto the planet and its CPD from the circumstellar disk \cite[e.g.,][]{Szula2014}. Regardless of the mechanism, one consequence of the accretion process is the presence of shocks on the planet surface and/or the CPD that can emit accretion tracers such as the H$_{\alpha}$ emission line.
Furthermore, several studies have suggested that giant accreting protoplanets could be detected using such tracers, particularly the H$_{\alpha}$ emission line \citep[e.g.,][]{Close2014,Zhu2015,Szu2020}.

\citet{Close2014} presented the first high-angular-resolution H$_{\alpha}$ imaging observations of a young star, HD142527, and reported the detection of a H$_{\alpha}$-emitting low-mass star embedded in its transitional disk, although the stellar or planetary nature of this latter source is a matter of ongoing discussion \citep{Brittain2020}.
Since then, several works have focused on  detecting accreting protoplanets through H$_{\alpha}$ imaging and/or spectro-astrometry, with the main targets being young stars surrounded by transitional disks \citep[see e.g.,][]{Sallum2015,Whelan2015,Huelamo2018,Mendigutia2018,Cugno2019, Zurlo2020,Uyama2020,Xie2020}. 
Currently, there are two bona fide accreting protoplanets detected in the H$_{\alpha}$ line: PDS70\,b and c \citep{Wagner2018,Haffert2019}. 
In parallel, several works have presented simulations to derive crucial accretion parameters from H$_{\alpha}$ luminosities and line profiles \citep[e.g.,][]{Aoyama2018, Thana2019, Marleau2022}.

Here we present the main results from a project designed to detect young accreting protoplanets within the gaps of five (pre-)transitional disks with the Spectro-Polarimetric High-contrast Exoplanet REsearch (SPHERE) and the Zurich Imaging Polarimeter (ZIMPOL) at the Very Large 
Telescope (VLT), which complements previous works by \citet{Cugno2019} and \citet{Zurlo2020}. Section~\ref{targets} provides information about the five observed targets, while Sections~\ref{data} and \ref{analysis} describe the data reduction and analysis. We present the main results and conclusions in the last two sections.

\section{The targets}\label{targets}

For this project, we selected young stars displaying structures in their disks, mainly gaps, as signposts of planet formation. We required that most of the gaps be spatially resolved by  the instrument SPHERE/ZIMPOL (see Section~\ref{data}), and that the central objects be sufficiently bright  ($R \lesssim$  12\,mag) 
to close the loop with Sphere AO for eXoplanet Observation (SAXO), the adaptive optics (AO) module of SPHERE. 

As explained above, the selected targets complement previous surveys performed with SPHERE/ZIMPOL:  \citet{Cugno2019} presented observations of six stars, four of them being intermediate-mass stars (spectral types between B9 and F6) observed as part of the Guaranteed Time Observation (GTO) program of the SPHERE consortium. 
These authors also included  data for TW~Hya (program 096.C-0267B , PI. Hu\'elamo) obtained under poor atmospheric conditions, and from the star  LkCa15 obtained as part of the ``Science Verification'' of the instrument.
Finally, \citet{Cugno2019} reanalyzed the data from the A8-type star MWC758 presented by \citet{Huelamo2018}.  Also, \citet{Zurlo2020} presented data for 11 additional transitional disks within 200\,pc and observable with SPHERE.

Here, we present observations of  four T~Tauri stars (RXJ1615, LkCa15, TW~Hya, T~Cha) and a Herbig Ae/Be star (HD\,163296). As explained above, LkCa15 (Science Verification, P.I. Hu\' elamo) and TW~Hya were previously observed by SPHERE, but under poor atmospheric conditions.
The main properties of the stellar sample are summarized in Table~\ref{stars1}. We briefly describe the targets in the following subsections. We note that we have updated all the angular separations provided using the GAIA EDR3 distances \citep{Gaia2016,GaiaEDR32021} included in  Table~\ref{stars1}.

\begin{table*}
\caption{Properties of the observed sample}\label{stars1}
\begin{tabular}{lccccccll}\hline
Target           & $R$ & $A_{\rm V}$ & $T_{\rm eff}$ &  distance$^1$ & $M_{*}$ &  $L_{*}$ & $R_{*}$ & Age \\
                    & (mag)  &  (mag)          & (K)  & [pc]                  & ($M_{\odot}$) & ($L_{\odot}$) & ($R_{\odot}$) & (Myr) \\ \hline
 RXJ1615      & 11.2     &  0.6$^2$     & 4000$^2$ & 155.6$\pm$0.6   & 0.6$\pm$0.1$^3$  & 0.90$\pm$0.02$^3$ & 2.0$^{\dagger}$  & 1.0$^{3}$ \\
 LkCa15       & 11.6     &  0.6$^2$     & 4800$^2$ & 157.2$\pm$0.6   & 1.2$\pm$0.1$^3$ & 1.11$\pm$0.04$^3$ & 1.5$^{\dagger}$ & 6.3$^{3}$\\ 
 TW~Hya       & 10.6     &  0.0$^2$     & 4000$^2$  & 60.1$\pm$0.1  & 0.8$\pm$0.1$^{3}$   &0.33$\pm$0.2$^{3}$ & 1.2$^{\dagger}$ & 6.3$^{3}$ \\
 T~Cha        & 10.4     &   1.2$^4$    & 5400$^5$ & 102.7$\pm$0.3   & 1.5$\pm$0.2$^{5,6}$ & 2.5$^5$ & 1.8$^{\dagger}$ &  5$^{+3}_{-2}$ $^7$ \\ 
 HD\,163296   &  6.9     &   0.0$^8$   & 8750$^8$ & 100.9$\pm$0.4   & 1.91$^{+0.12}_{-0.0}$ $^8$         &  15.5$^{+1.5}_{-}$ $^8$ & 1.70$^8$ & 10$^8$ \\ \hline
\end{tabular}

Note: $^{1}$ GAIA EDR3 distances \citep{GaiaEDR32021};  $^2$ \citet{Garufi2018}; $^3$ Adopted from \citet{ATorres2021}; $^4$ \citet{Cahill2019}; $^5$ \citet{Schisano2009}, estimated for a distance of 100\,pc; $^6$ \citet{Huelamo2015}; $^7$ \citet{Dickson2021};  $^8$ \citet{GuzmanDiaz2021}; $^{\dagger}$ Estimated from $T_{\rm eff}$ and $L_{*}$.

\end{table*}

\begin{table*}\label{obslog}
\caption{Observing log}
\begin{tabular}{lcccrcc}\hline
Target           & Obs. Date & DIT     & Total Exptime & Field Rotation & Airmass$^{*}$   & DIMM $\tau_0^*$ \\ 
                 &            & [sec]  & [sec] & [degrees]        &                 & [ms]  \\ \hline 
 RXJ1615          & 2018 Apr 20   & 60  & 7200 & 123.2 & 1.02 & 8 \\ 
 LkCa15           & 2018 Dec 19  & 80 & 6720 & 33.9    & 1.48 & 4 \\ 
 TW~Hya           & 2019 Mar 15  & 80  & 6720 & 107.6 & 1.02 &  5 \\ 
 T~Cha            &  2019 Mar 22  & 70 & 6300 & 29.3   & 1.74 & 9 \\
 HD\,163296$^{\dagger}$      & 2018 Aug 18  &  20 & 4400 & 7.8     & 1.06 & 7  \\ 
 HD\,163296      &  2018 Sep 06 & 20  & 5100 & 8.9     & 1.15 & 6 \\
\hline
 \end{tabular}

Note: $^{*}$Average value along the exposure. 
\end{table*}
\subsection{RXJ1615.3-3255}

RXJ1615.3-3255 (RXJ1615 hereafter) is a member of the Lupus star forming region, and was identified as a T Tauri star (TTS) in a X-ray survey with the ROSAT satellite \citep{Krautter1997}. 
Analysis of {\em Spitzer} spectroscopy suggested that RXJ1615 was surrounded by a transitional disk \citep{Merin2010}, which was spatially resolved at 880\,$\mu$m by the Submillimeter Array (SMA) revealing a dust cavity of $\sim$25\,au in size, and a disk inclination of $\sim$41$\degree$  \citep{Andrews2011}.
Later, \citet{Vandermarel2015} presented resolved images of the disk using ALMA Band 9 observations (440\,$\mu$m). These authors reported a smaller dust cavity ($\sim$17\,au), and the  presence of a second dust cavity in the outer part of the disk, between $\sim$ 0\farcs6 and 0\farcs7 (93-109\,au).

\citet{deBoer2016} obtained polarimetric observations of the target with VLT/SPHERE in the optical and near-infrared (NIR) regime, which were sensitive to small dust particles in the disk surface. The images revealed a complex disk architecture with the presence of two arcs, two rings, an outer gap at $\sim$0\farcs5 ($\sim$78\,au), and an inner disk. These authors only reported marginal evidence for an inner cavity of 25 au. They also looked for close companions, detecting nine objects between two and eight arcseconds, outside the disk. 
Four of these were confirmed as background sources, while the other five require additional data in order to confirm or reject them as co-moving companions. 
\citet{Willson2016} also looked for close companions around RXJ1615 through sparse aperture masking $K$-band observations. Although these latter authors  detected a significant asymmetry in the closure phases, the identification of systematic effects on their data prevented them from drawing firm conclusions about the presence of potential companions.

\citet{Avenhaus2018} presented SPHERE/IRDIS polarimetric differential imaging of the disk in the $J$ and $H$ bands. The results obtained by these authors suggested a smaller inner cavity than that reported by \citet{Vandermarel2015}, as they detected  scattered light down to $\sim$0\farcs1 (15.6\,au), which corresponds to the edge of the used coronograph.
By modeling the inner cavity, \citet{ATorres2021} estimated that a planet with $\sim$4.5\,M$_{Jup}$ at 22\,au might have carved the observed gap, but they could not detect it.

\subsection{LkCa~15}

LkCa15 is surrounded by a pre-transitional disk with an inner disk, a large gap with a minimum at $\sim$43\,au, and an outer disk at 58\,au \citep{Thalmann2016}.
This TTS was studied by \citet{Kraus2012} through SAM interferometry in the NIR,  reporting the detection of a planet candidate in the gap of the disk (LkCa15b). \citet{Sallum2015}  confirmed the detection of this IR signal with new SAM IR data. These authors also detected two additional IR point-like sources (LkCa15\,c and d), and the three detections were interpreted as protoplanet candidates. One of them, LkCa15\,b, was also detected in H$_{\alpha}$ imaging observations at a separation of 93\,mas ($\sim$ 15\,au) and PA of 256\,degrees. 
However, \citet{Mendigutia2018} did not detect such a planet using spectro-astrometry, instead suggesting that the H$_{\alpha}$ emission is extended and roughly symmetric. 
In fact, \citet{Thalmann2016} and \citet{Currie2019} questioned the existence of these protoplanets, proposing that previous IR detections are in fact related with clumpy emission from the inner disk. 

Using very high-angular-resolution ALMA continuum observations at 1.3\,mm,  \citet{Facchini2020} estimated that the flux detected in the innermost regions of LkCa15 could be related to an inner disk not larger than 0.15\,au in radius, assuming optically thick dust emission. \citet{Close2020} explained that this result might indicate that the bright clumps detected at IR wavelengths could be related to circumplanetary disks and not the inner disk. In addition, this latter author remarks that the reported H$_{\alpha}$ emission cannot be explained by any mechanism other than a young accreting protoplanet.
In this context, an observation at a  second epoch confirming the H$_{\alpha}$ detection of LkCa15b would be extremely useful.

\subsection{TW~Hya}

TW~Hya has been extensively studied using NIR and optical imaging, and spectroscopic observations searching for planet companions \citep[e.g.,][]{Huelamo2008,Biller2013,Ruane2017,Uyama2017}, without any confirmed detection. Several works have analyzed the face-on disk around TW Hya in optical, IR, and submillimeter(submm) light:
\citet{Andrews2016} presented very high-angular-resolution 870$\mu$m ALMA observations of its disk, revealing an inner gap at $\sim$1\,au, and three dark annuli at 25, 42, and 49\,au. 
\citet{vanBoekel2017} presented SPHERE optical and IR polarimetric observations of the source, revealing two clear gaps at 23\,au and 94\,au, and a tentative third gap at $<$ 7\,au. These authors estimated that the mass of potential protoplanet companions carving the gaps should be a few 10\,M$_{\Earth}$  at most.
\citet{Dong2017} estimated the masses of the perturbers causing the two outer gaps to be 0.15 and 0.08 M$_{\rm Jup}$, respectively, while \citet{ATorres2021} estimated a mass of $\sim$0.04\,M$_{\rm Jup}$ to explain the gap at 7\,au.

\citet{Tsu2019} reported a protoplanet candidate in TW~Hya based on high-resolution ALMA observations at 1.3\,mm. The candidate is detected at a separation of 0\farcs87 (52\,au) and a position angle of $\sim$237$^{\degree}$. They estimate that the mass of the candidate is close to a Neptune-mass planet.

 Finally, different works have looked for protoplanets inside the disk gaps using different accretion indicators (e.g., Br$_{\gamma }$,Pa$_{\beta}$, H$_{\alpha}$),  resulting in nondetections \citep[see][]{Uyama2017,Cugno2019}.

\subsection{T~Cha}

T~Cha is a young star with a highly inclined ($i\sim$ 67 degrees) transitional disk resolved at (sub)mm, mm, and optical wavelengths \citep{Huelamo2015,Pohl2017a,Hendler2018}.
A possible substellar companion candidate was reported by \citet{Huelamo2011}, but several authors \citep{Olofsson2013, Cheetham2015, Sallum2015} associated the NIR detection with the inner disk around the source. Spectro-astrometry observations in the H$_{\alpha}$ line did not reveal any accreting companion \citep{Cahill2019}. 

 Optical polarimetric observations of the target with SPHERE/VLT resolved a gap of 0\farcs28 ($\sim$29\,au) in size \citep{Pohl2017a}. Using ALMA at 3 and 1.6\,mm, \citet{Hendler2018} detected an unresolved inner disk at radius $<$ 1\,au, and resolved a disk gap with a width of between 17 and 27\,au, with the gap outer radius located at $\sim$22-27\,au. These authors conclude that the gap could have been created by either a single $\sim$1.2 M$_{\rm Jup}$ planet or by multiple (and less massive) planets.

\subsection{HD\,163296}\label{hd163296}

This is the only Herbig Ae/Be star of the sample.  It is an A1-type star at a distance of $\sim$101\,pc surrounded by a dusty disk with a radius of $\sim$ 200\,au, and a gaseous disk of more than twice this size \citep{deGregorioMonsalvo2013}. The star was imaged with ALMA at high angular resolution, revealing  at least three dust gaps with radii of $\sim$45, 87, and 140\,au \citep{Isella2016}. \citet{Liu2018} explained these three dust gaps with the presence of planets with masses of $\sim$0.5\,M$_{\rm Jup}$. New ALMA observations of HD163296 revealed an additional inner gap at $\sim$10\,au \citep{Isella2018}.

\citet{Teague2018} used the rotation curve of the CO molecule to infer the presence of two planets at 83 and 137\,au, with masses of $\sim$1 and 1.3 M$_{\rm Jup}$, respectively.  \citet{Pinte2018} also presented the detection of a large, localized deviation from Keplerian velocity in the proto-planetary disk. 
Their derived velocity pattern is consistent with the dynamical effect of a $\sim$2\,M$_{\rm Jup}$ planet orbiting at a radius of $\sim$2.4\,arcsec ($\sim$242\,au) from the star, that is, outside the dusty disk.  \citet{Pinte2020} identified a second planet candidate (with a mass of the order of 1 M$_{\rm Jup}$ in HD163296 at 0\farcs67, and position angle of -93 degrees.
\citet{Guidi2018}, presented Keck $L$-band data for the system, detecting a  6\,M$_{\rm Jup}$ candidate at 0.5 arcsec (50\,au) from the star and PA$\sim$30$^{\degree}$, close to the inner edge of the 45\,au gap.  These authors also derived planet mass limits of 8-15, 4.5-6.5, and 2.5-4\,M$_{\rm Jup}$ at the position of the three gaps reported by \citet{Isella2016}.  \citet{Mesa2019} presented SPHERE/IRDIS and IFS IR observations of HD163296, and did not detect any protoplanet, but put limits on the mass of potential planet companions by comparing their contrast curves with pre-main sequence evolutionary models. 
We note that, very recently, \citet{Izquierdo2021} identified a new planet candidate (projected separation of 0.77\,arscec, and PA of $\sim$352$^{\degree}$) 
through kinematic analysis of ALMA data.


\section{SPHERE/ZIMPOL observations and data reduction}\label{data}

The observations presented here were obtained in service mode with SPHERE/ZIMPOL in 2018 and 2019. SPHERE \citep{beuzit2008} is an extreme adaptive optics (AO) system for high-resolution and high-contrast observations at the VLT, and ZIMPOL \citep{Schmid2018} is the Imager and Polarimeter instrument operating at optical wavelengths.
ZIMPOL was used in spectral and angular differential imaging modes \citep{marois2006,racine1999} with the dichroic beam splitter, so that all the targets were imaged simultaneously in two different filters, N\_Ha\footnote{There are typos in the Ha-filter wavelengths in \citet{Schmid2018}. Correct values are provided in \citet{Schmid2017} (Table~5).}
($\lambda_{c}$ = 656.5\,nm, $\Delta \lambda$ = 0.97\,nm) in the filter wheel~2 (FW2) and Cnt\_Ha 
($\lambda_{c}$ = 644.9\,nm, $\Delta \lambda$ = 4.1\,nm) in the filter wheel~1 (FW1), in pupil stabilized mode. The total field of view covered by the ZIMPOL detector is of $\sim$ 3\farcs6 $\times$ 3\farcs6, with a pixel scale of 3.6\,mas/pixel. The detector gain is 10.5e-/ADU with a readout noise of 20 e-/pixel. We use SAXO \citep{Petit2014} 
to close the loop on the central stars, which have R-mag values of between 6.9 and 11.6\,mag.  The details of the observations for each target are included in Table~\ref{obslog}. 

\begin{figure}[t!]
\includegraphics[width=8.5cm]{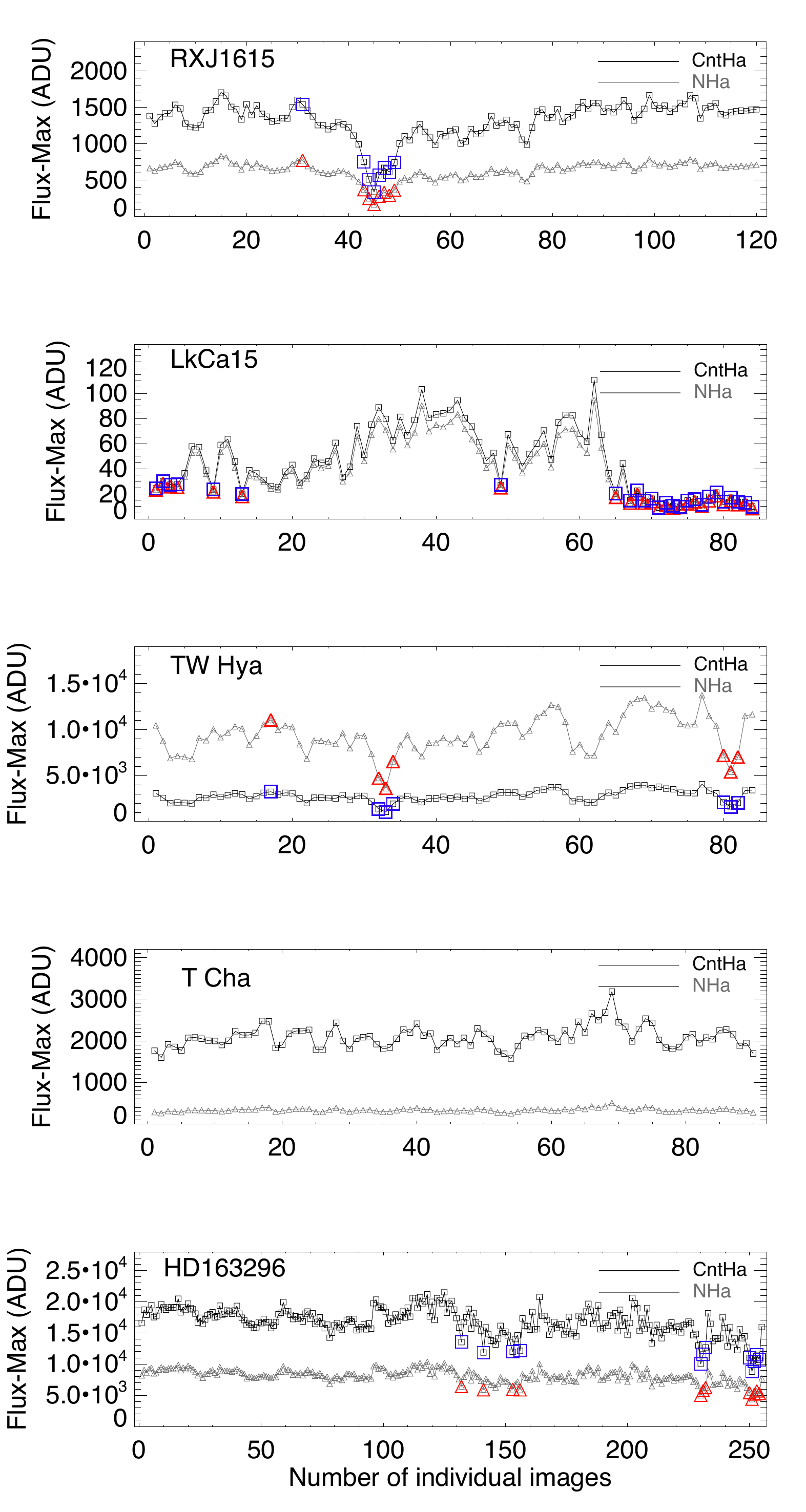} 
\caption{Maximum counts (in ADU) registered in the two filters along the whole exposure for the five targets. The blue squares in the CntHa filter (or red triangles in the NHa one) depict the rejected images in each dataset.}\label{aduflux}
 \end{figure}

 All the targets were observed under clear conditions. However, for some objects the atmospheric conditions were variable. To study the impact of this variability on the AO correction, we present the maximum counts registered on the targets in the two filters (this is a good proxy for the Strehl variation in photometric or clear conditions)  along the exposures  in Figure~\ref{aduflux}.  For each target, we discarded the individual images with the poorest AO correction (displaying a significant decrease in the maximum counts in both filters along the observing sequence) or those miscentered, which are shown in blue (Cnt\_Ha) and red (N\_Ha) points in Fig.~\ref{aduflux}. Table~\ref{obslog} also includes the average value of the coherence time ($\tau_o$) for each dataset measured at the DIMM monitor at $\sim$500\,nm.
 
The data reduction was performed using a ZIMPOL customary pipeline developed at ETH Z\"urich. As a first step, the images were remapped to a square grid of 1024$\times$1024 pix$^2$ with a pixel scale of 3\farcs6$\times$3\farcs6, bias-subtracted, and flat-fielded. Subsequently, the individual images were recentered using a simple Moffat function to estimate the centroid position, as no coronograph was used. This resulted in two individual cubes, one per filter.
For the point-spread function (PSF) subtraction, the individual filter datasets (N\_Ha and Cnt\_Ha) were considered separately to apply a standard angular differential imaging (ADI) processing technique using classical-ADI (cADI), smart-ADI (sADI), and radial-ADI 
\citep[rADI, see][for a description of these techniques]{Lafreniere2007,Chauvin2012}, and  (smart) principal component analysis (PCA and sPCA) with one, five, and ten components \citep{Soummer2012}. We note that sPCA is a simple variation of the PCA approach that minimizes the self-subtraction of the signal by rejecting frames too close in time for the eigen-modes calculation. The purpose of using all these algorithms, which handle the speckle noise in different ways, is to secure the detection of point-like sources in the innermost regions of the images. 

  Simultaneous observations with ZIMPOL ensure that images in two filters present the same speckle pattern. To exploit the spectral diversity of the two filters (Cnt\_Ha and N\_Ha, inside and outside the H$_{\alpha}$ line, respectively),
  we applied a dedicated angular and spectral differential imaging (ASDI) process.
  The individual Cnt\_Ha images were first spatially rescaled to the N\_Ha filter resolution, and then flux-normalized considering the total flux ratio between N\_Ha and Cnt\_Ha within an aperture of $r=10$ pixels. Finally, we performed the subtraction N\_Ha-Cnt\_Ha frame per frame to exploit the simultaneity of the two observations. ADI processing in cADI, sADI, PCA, and sPCA was then applied to the resulting differential datacube.
  An example of all the processing algorithms applied to one of the targets (TW Hya) is shown in Figure~\ref{twhya_all}.
  
\begin{figure*}[p]
 \center
\includegraphics[width=4.6cm,angle=270]{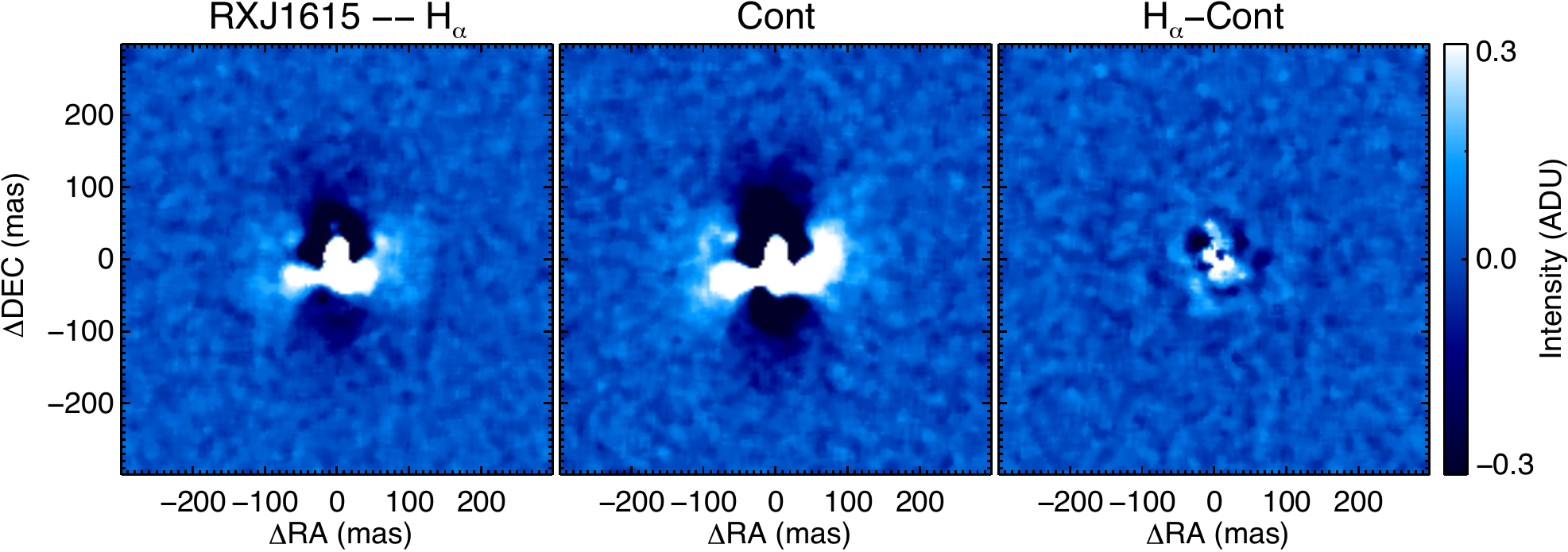}
 \includegraphics[width=4.6cm,angle=270]{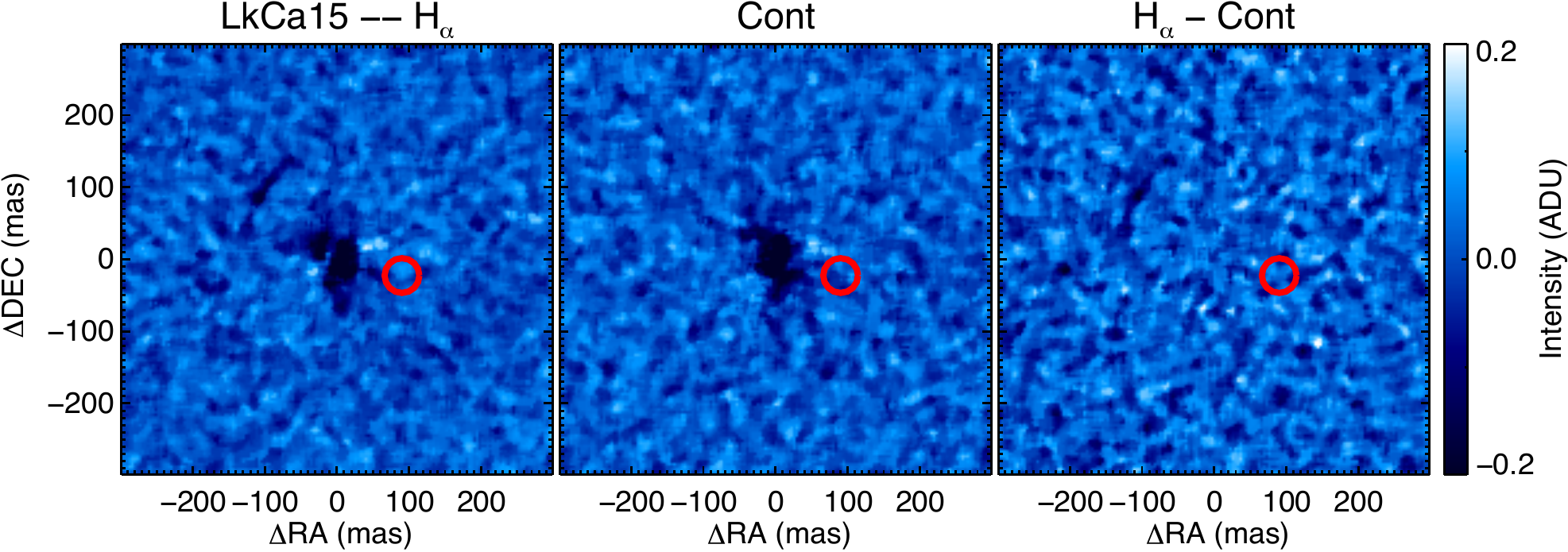}
\includegraphics[width=4.6cm,angle=270]{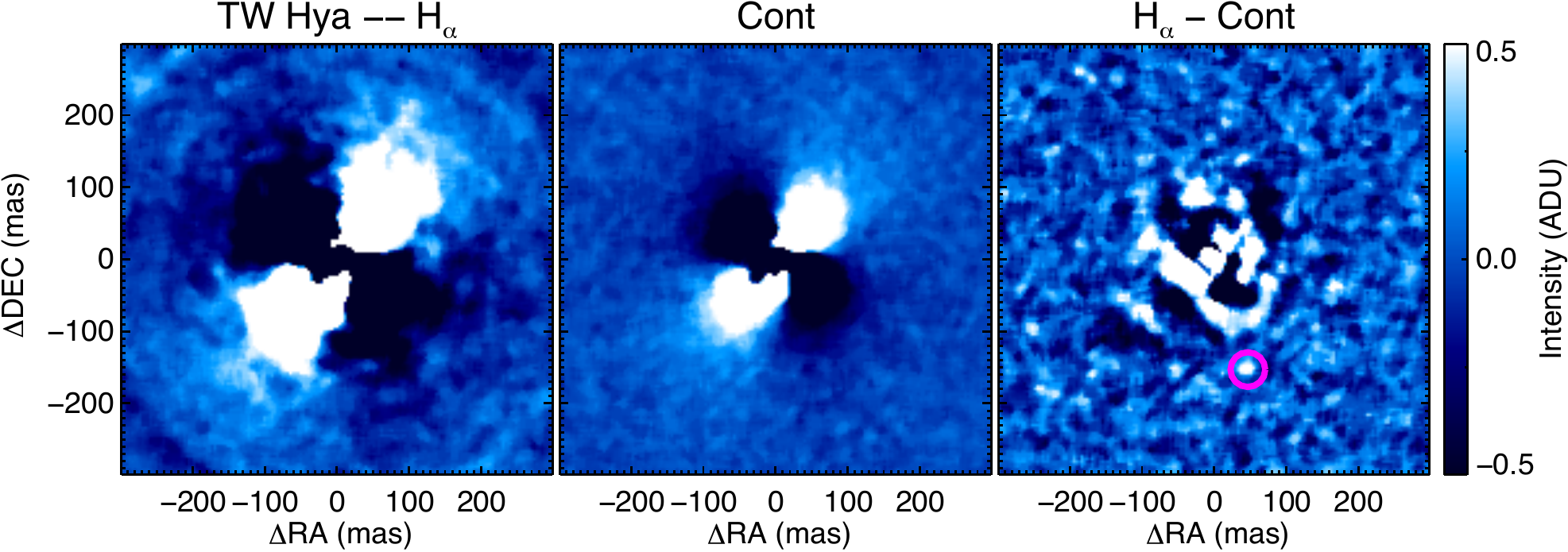}
  \includegraphics[width=4.6cm,angle=270]{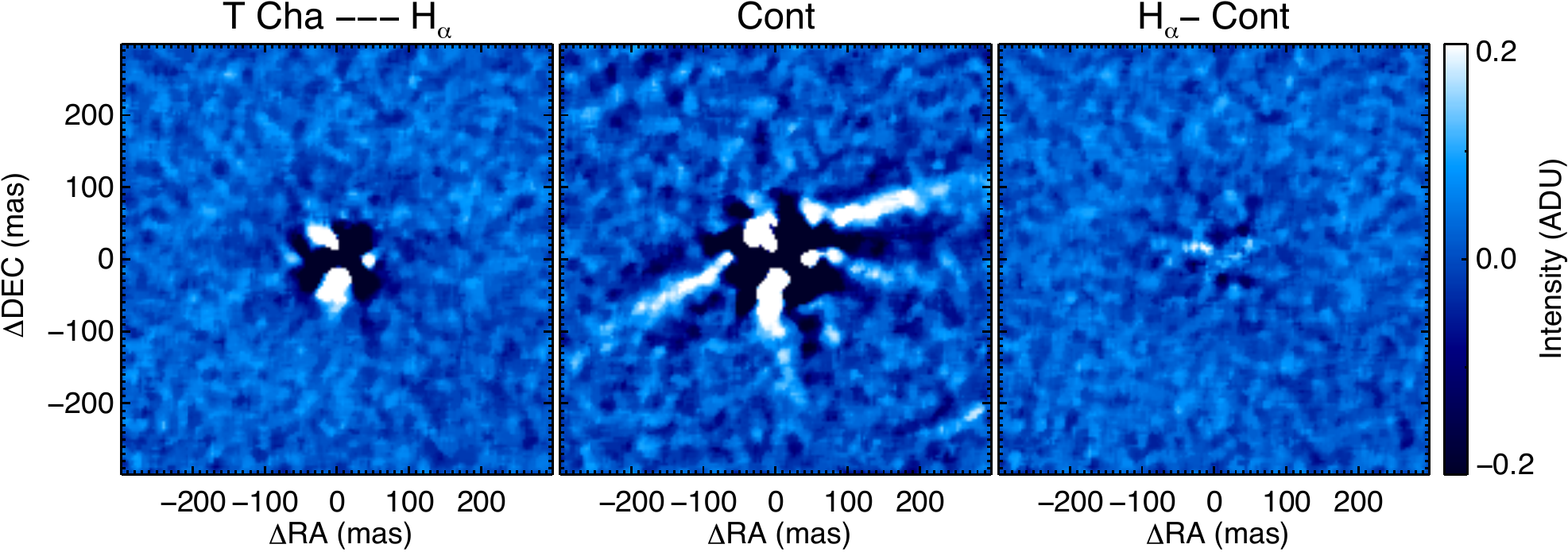}  
 \includegraphics[width=4.6cm,angle=270]{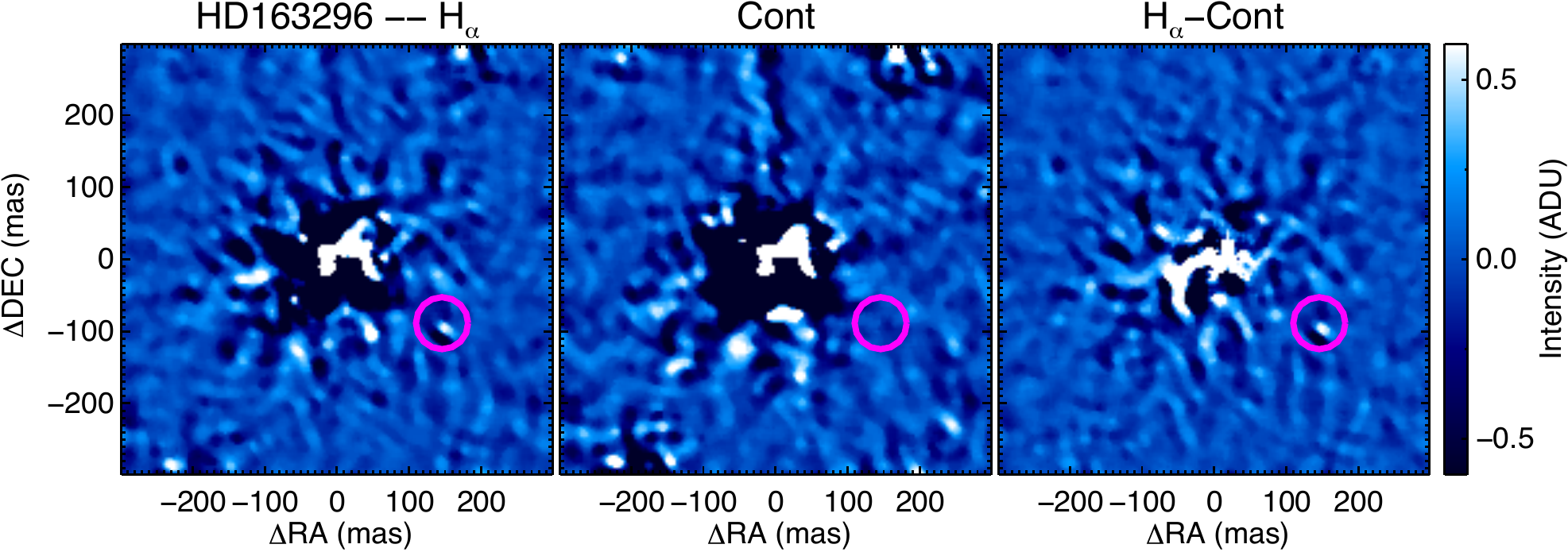}
  \caption{Final cADI images of the five stars observed with SPHERE/ZIMPOL. The panels show the H$_{\alpha}$ (left) and continuum (middle) images, and their difference (ASDI image, right). All the panels show a FOV of $\sim$ 0.6"$\times$0.6". In the case of LkCa15, we mark the position of the candidate LkCA15b with a red open circle, while in the case of TW~Hya and HD163296 we mark  point-like faint detections with magenta circles (see text).}\label{imagesCADI}
 \end{figure*}

  Figure~\ref{imagesCADI} shows the final cADI N\_Ha, Cnt\_Ha and N\_Ha-Cnt\_Ha (ASDI) images of all the observed targets.
We note that all the images have been rotated 134$^{\degree}$ counterclockwise to align the camera with the north upwards and east to the left \citep[see][]{Cugno2019}.

%

\begin{figure}[t]  
\includegraphics[width=0.5\textwidth]{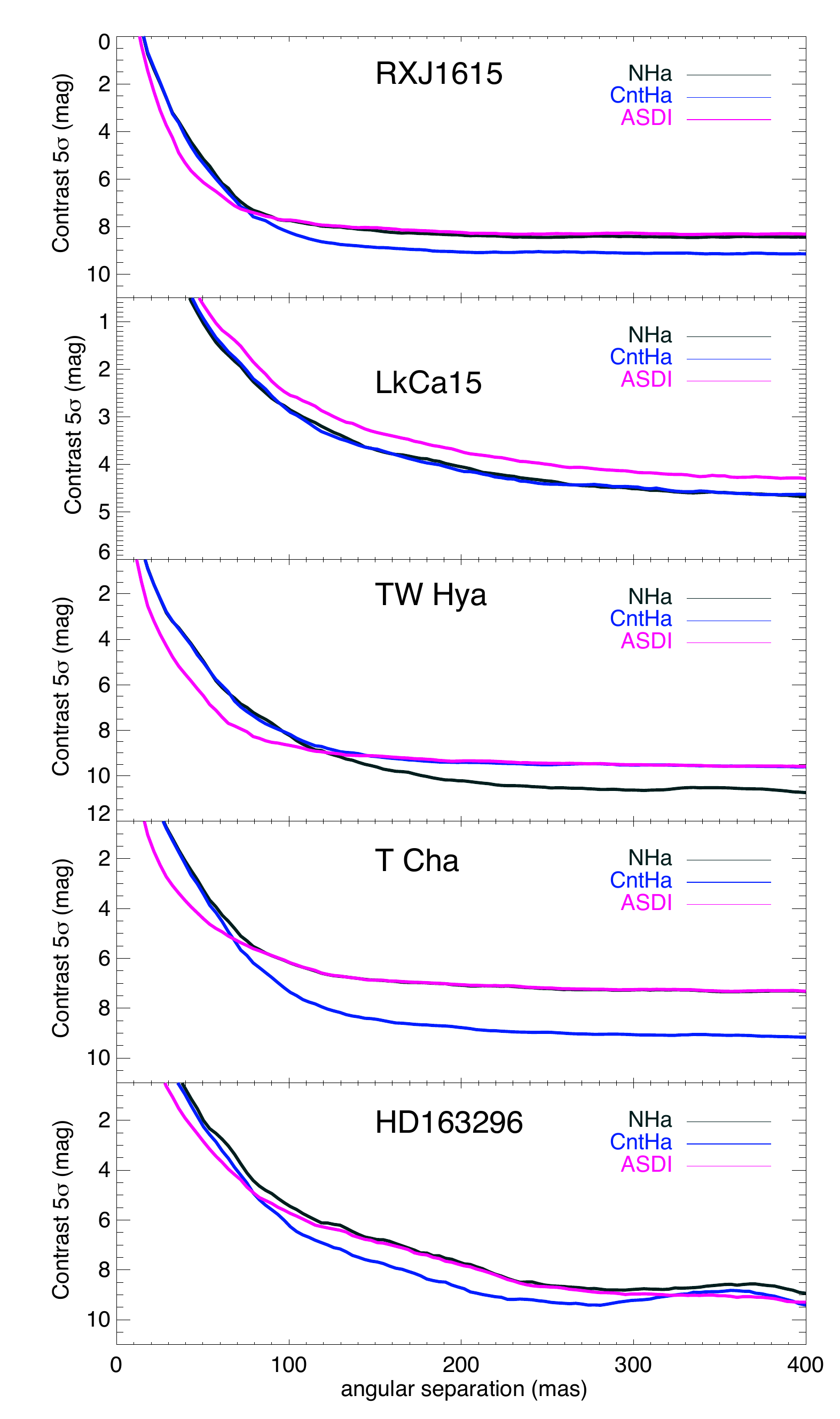}
 \caption{Contrast curves derived from the H$_{\alpha}$, continuum, and ASDI (N\_Ha - Cnt\_Ha) reduced images of the five observed objects.}\label{limdet}
 \end{figure}

The ADI detection limits in the two filters were estimated from a standard pixel-to-pixel noise map of each filter within a box of 5$\times$5 pixels sliding from the star to the limit of the ZIMPOL field of view. 
To correct for the flux loss related to the ADI processing, fake planets were regularly injected into the original data cubes before PSF subtraction (every 20 pixels at three
different position angles, with a flux corresponding to
100\,ADU). The cubes were then reprocessed using the different ADI algorithms, allowing us to estimate the flux loss by computing
the azimuthal average of the flux losses for fake planets at the same radii \citep[e.g.,][]{Jorquera2021}.
The contrast curves at 5$\sigma$ were then obtained using the pixel-to-pixel noise map divided by the ADI flux loss and corrected for small number statistics following the prescription of \citet{Mawet2014} but adapted to our 5$\sigma$ confidence level at small angles with ZIMPOL. The same method was applied for the ASDI dataset to illustrate the gain in speckle subtraction in the inner region below $\sim$100\,mas. All the contrast curves  are displayed in Figure~\ref{limdet}. 
We note that the Strehl variations along the exposures (that can be of 20\%-30\% according to Fig.~\ref{aduflux}) can affect the
PSF normalization that we apply to build the contrast curves, 
which are therefore affected by an error bar of $\sim$ 0.2-0.3\,mag. This does not apply to LkCa15, in which the Strehl variations are significantly larger, and therefore 
this source is expected to show higher uncertainties.

\section{Data analysis}\label{analysis}

As a general result, we do not detect any obvious point-like companion in the analyzed datasets. In the cases of TW~Hya and HD163296, we detect some faint point-like emission, the nature of which is discussed in the corresponding subsections below.

Our analysis of the data reveals that, for the cases of TW Hya, T Cha, and RXJ1615, the best contrast in the ADI Cnt\_Ha and N\_Ha images is obtained for the sADI and sPCA reductions at small separations, while cADI is better at larger separations given the larger number of frames used, minimizing the read-out and background noise.
In the case of LkCa15, the faintest object of the sample, cADI provides the best contrast even at small separations, as we are probably entering a low-flux regime for which the temporal evolution of the PSF becomes less critical. For all these objects, the best contrast in the ASDI images is obtained with the cADI reduction: for these images, the SDI technique is very efficient at suppressing temporally variable speckles in the individual N\_Ha-Cnt\_Ha images, meaning that the subtraction of a median PSF in the final image (cADI reduction) results in a high contrast. Finally, the case of HD\,163296 is different, because the small rotation field results in an important self-subtraction in the final images. For this object, we then analyzed the rotated and stacked image without subtracting the PSF, the so-called nADI image.  In the following subsections, we discuss the main results for each individual target.

\subsection{RXJ1615}

RXJ1615 was observed under good and stable conditions. We only rejected seven individual images,  mainly because of a seeing variation during the exposure (changing from 0.6 to 1.0\,arcsec, according to the guide probe). The ADI N\_Ha and the Cnt\_Ha images displayed in Figure~\ref{imagesCADI} show residuals that are not present in the ASDI image, which shows the best contrast at separations below 80 mas.

\subsection{LkCa~15}

LkCa~15 was observed in December 2018 for  $\sim$1.9h, covering a total field rotation of  34\,degrees. The average coherence time during the exposure was of $\sim$4 ms. We note that with an R-mag $\sim$12\, mag and a declination of $\sim$+24 deg, this is a challenging target for observation with SPHERE/ZIMPOL from the southern hemisphere.
The analysis of the maximum counts from the central object reveals very variable conditions, and a significant decrease in the flux in the second part of the exposure with a clear impact on the AO correction quality.  Indeed, the individual images show a very variable and elongated PSF.
To generate the final images, we selected only those data with the largest number of counts, resulting in a final exposure time of $\sim$1.3\,hours and a field rotation of $\sim$25\,degrees. %

As can be seen in Fig.~\ref{imagesCADI}, we do not detect any object in the final images.  We improve the contrast curve presented by \citet{Cugno2019} but, given the faintness of the target and the variable conditions, we do not reach sufficient sensitivity to detect the reported companion candidate (Fig.~\ref{limdet}). We reach a contrast of $\sim$2.5\,mag at the position of LkCa15b reported by \citet{Sallum2015},
which is far from the contrast of 5.2\,mag achieved with MagAO.

\subsection{TW~Hya}\label{sub_twhya}

For TW Hya, the atmospheric conditions were very good and stable, and so we only rejected 7 images (out of 82).  The total rotation field is $\sim$108\,degrees. The images in the two individual filters show an elongated PSF, and the elongation direction changes with the parallactic angle along the exposure. As a result, the final images show a ``butterfly'' pattern. This effect, known as a wind-driven halo (WDH),  was studied in detail by \citet{Cantalloube2020} and is related to high wind-speeds in the upper layers of the atmosphere.
Thanks to the simultaneity of the N\_Ha and Cnt\_Ha images, this elongated emission is partially suppressed in the ASDI image.
 
 The ASDI image in Fig.~\ref{imagesCADI} shows a faint emission feature located at a separation of $\sim$160\,mas SW (magenta circle). We note that this is the only faint emission that is detected in all of the ASDI image regardless of the data processing applied (e.g., cADI, sADI,rADI, PCA or sPCA; see Figure~\ref{twhya_all} in the Appendix). Inspection of the PSF in the individual N\_Ha and CntHa images along the exposure reveals some pinned speckles associated with the spiders that change in flux during the observation. Therefore, it is highly probable that the point-like emission observed in the ASDI images is caused by one of these speckles. Finally, we do not detect any source at the position of the ALMA candidate reported by \citet{Tsu2019}.

\subsection{T~Cha}

T~Cha was observed under good and stable conditions, and we did not reject any individual image.  The final images do not show any point source, and the contrast in the innermost regions reaches values of 5\,mag at 70\,mas in the ASDI image. The continuum image shows some radially extended structures. These resemble those detected at IR wavelengths by \citet{Pohl2017a}, and might be related to the disk itself. We note that this is a difficult object for the detection of accreting protoplanets, given the high disk inclination \citep[$\sim$ 69 degrees,][]{Huelamo2015}.

\begin{figure}[t!]
 \center
\includegraphics[width=0.2\textwidth,angle=270]{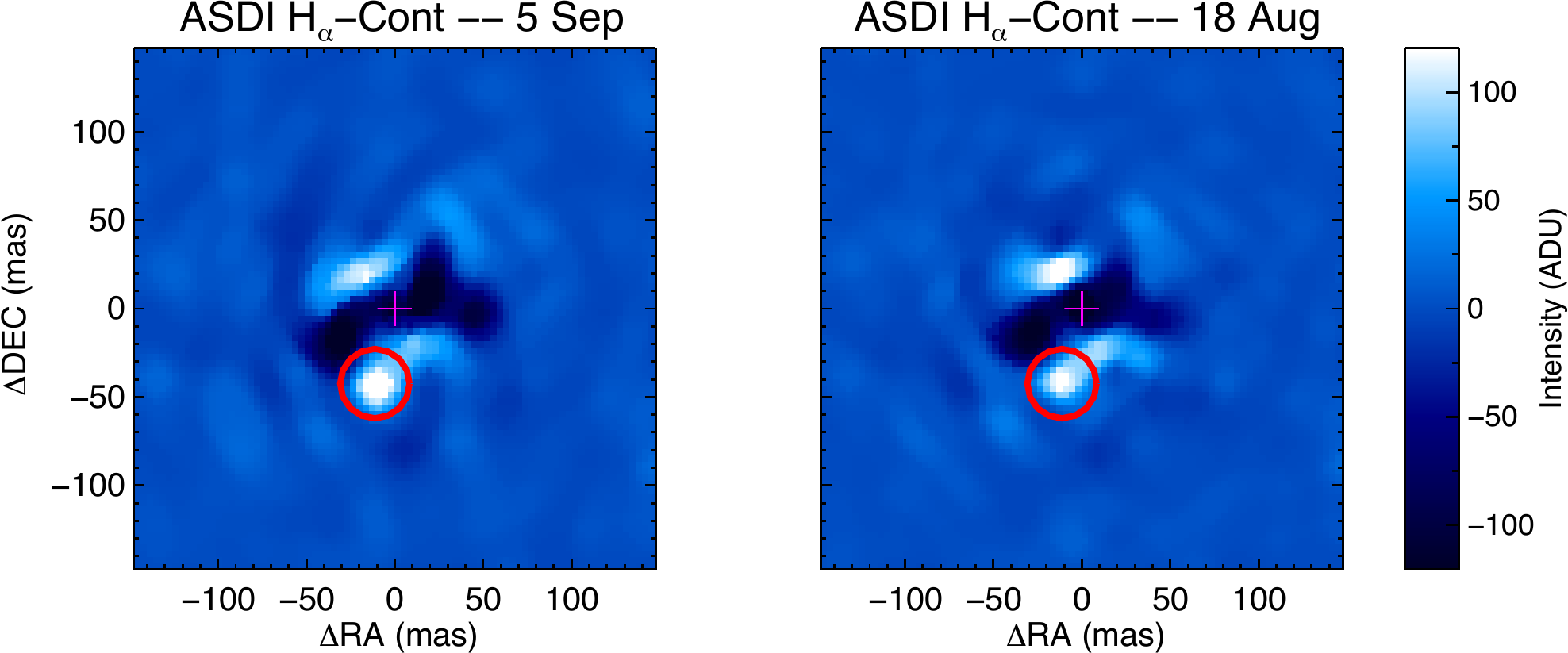}
\includegraphics[width=0.2\textwidth,angle=270]{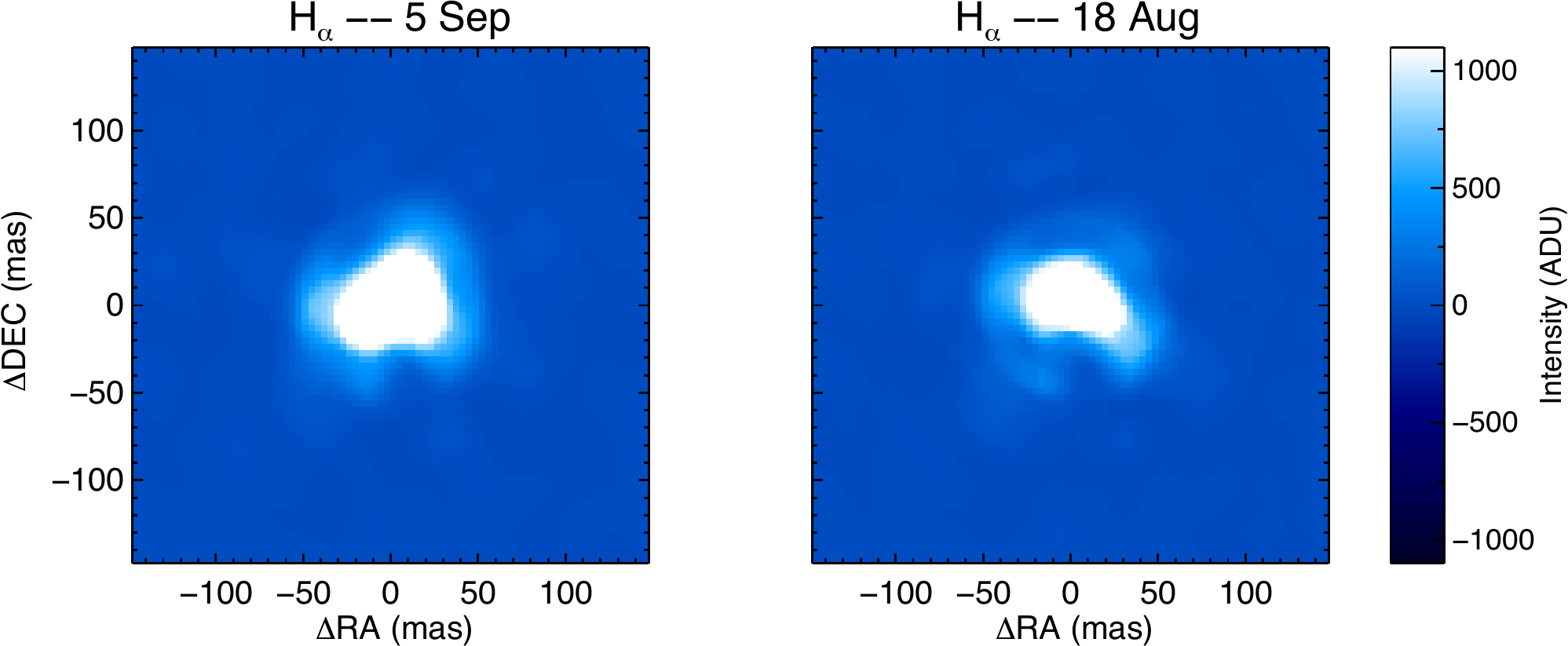}
\includegraphics[width=0.2\textwidth,angle=270]{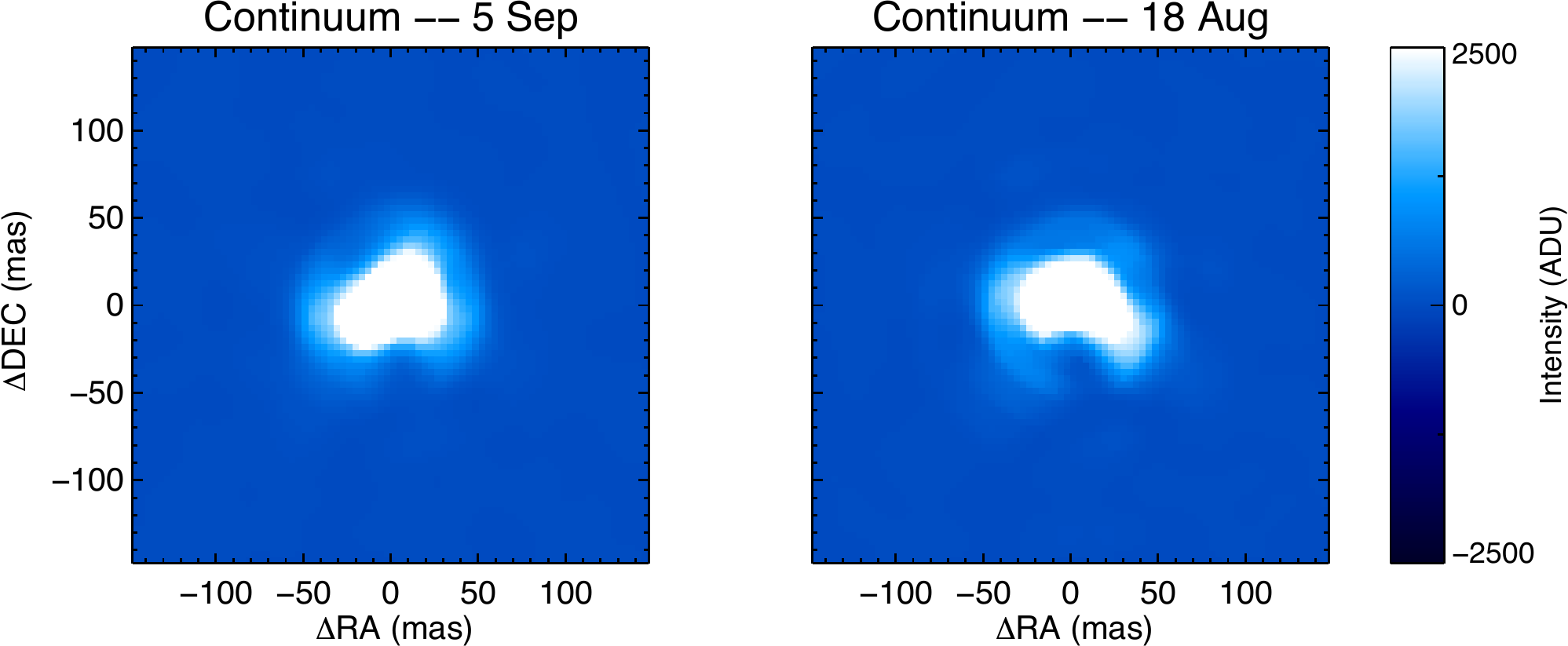}
\caption{nADI images of HD163296: the nADI reduction includes stacked images after derotation but without subtracting the PSF. The left panels include the  reduced data from September 5, while the right panels include data obtained on August 18 for comparison. From top to bottom: ASDI, N\_Ha and Cont\_Ha final images. The top panels display the position of the star (magenta cross) and the detected point-like emission (red circle) at $\sim$43\,au. We display a FOV of 0\farcs29$\times$0\farcs29. 
}\label{hd1632_nadi}
\end{figure}

\begin{figure}[t!]
 \center
\includegraphics[width=0.2\textwidth,angle=270]{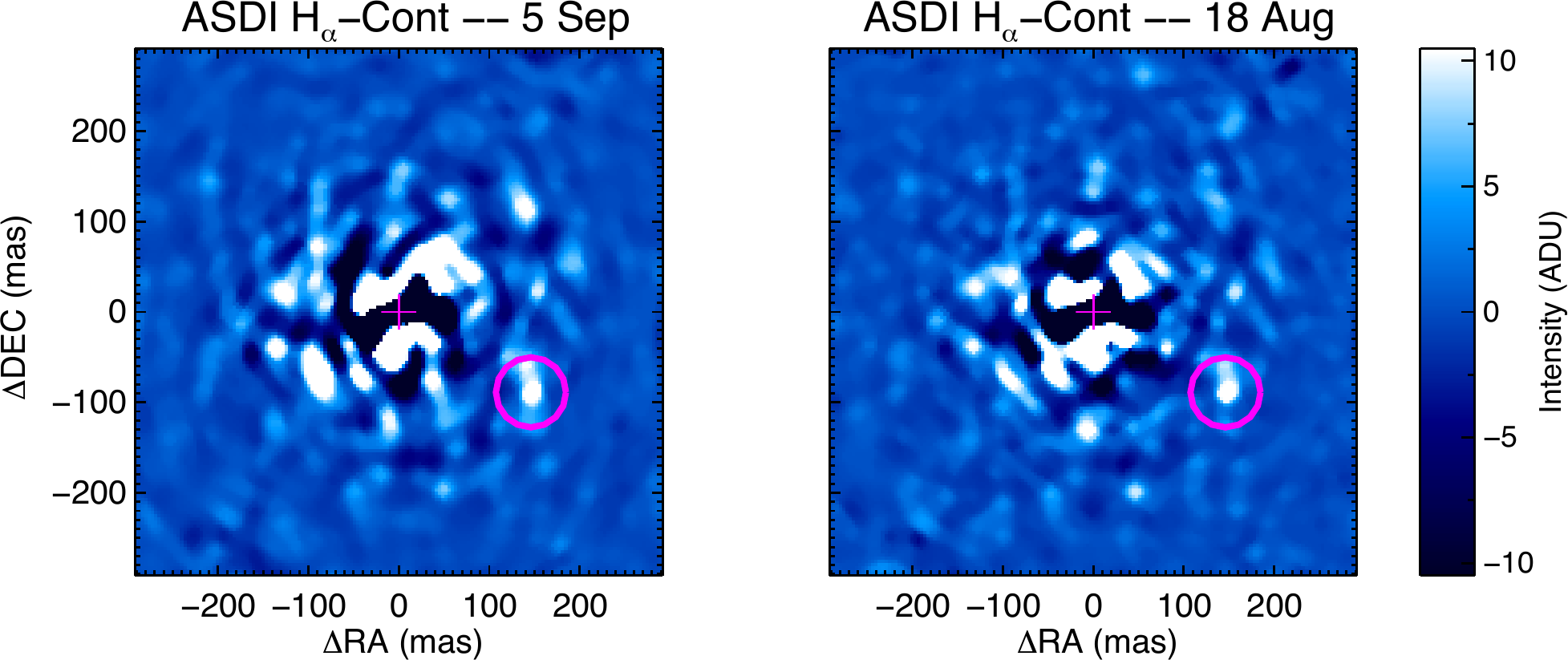}
\includegraphics[width=0.2\textwidth,angle=270]{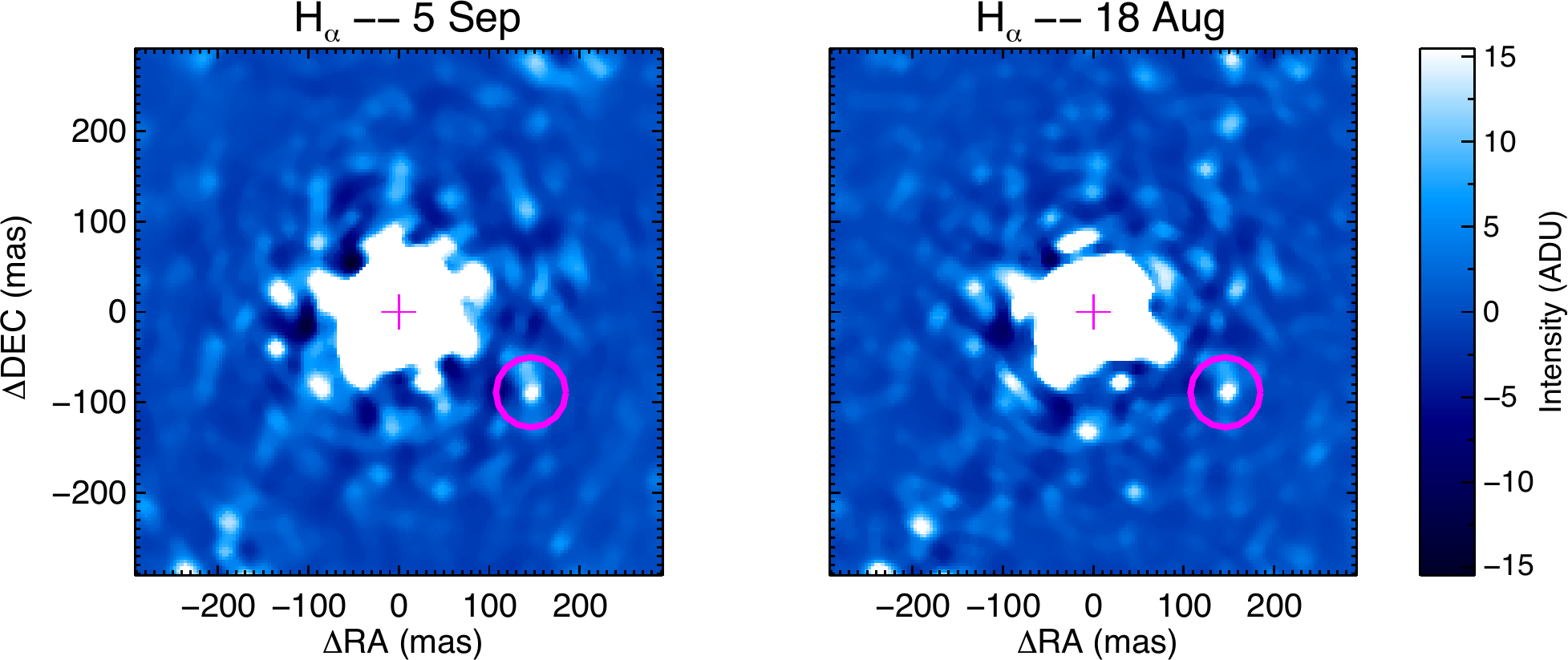}
\includegraphics[width=0.2\textwidth,angle=270]{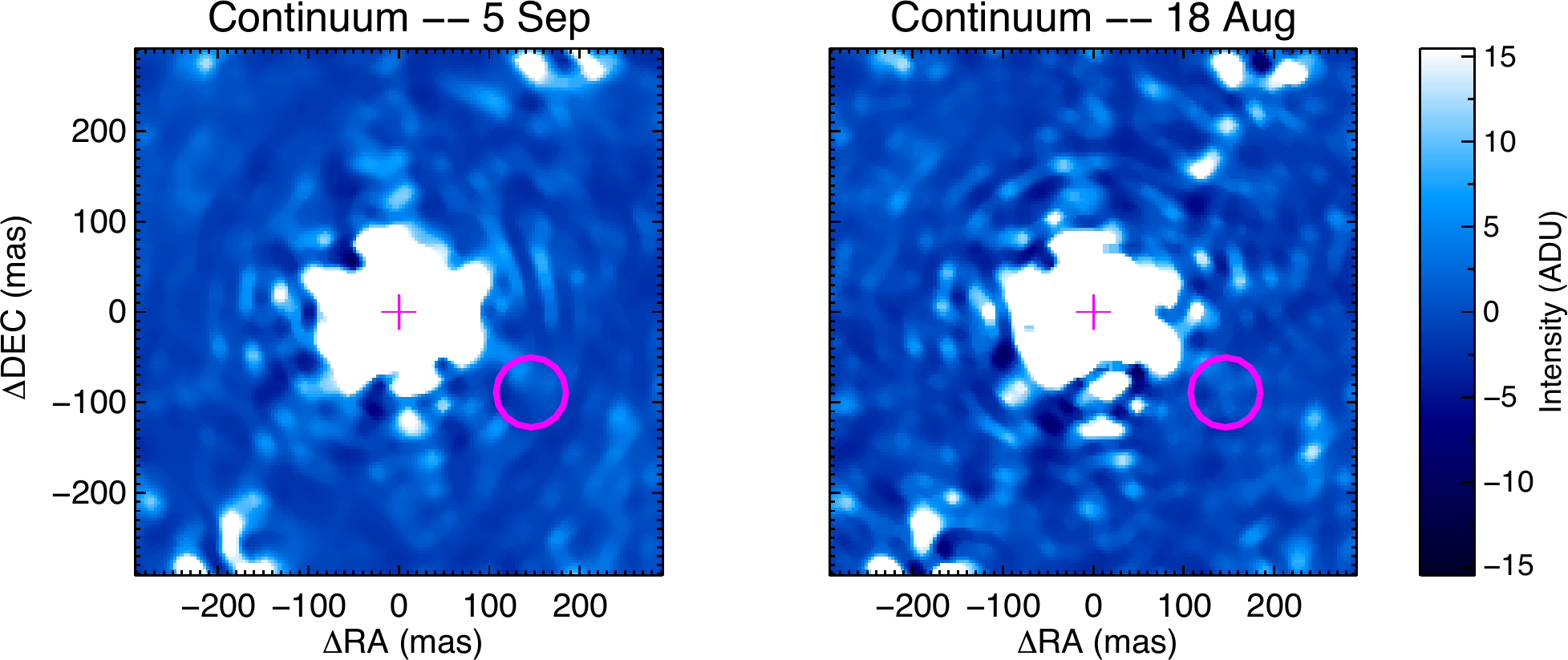}
\caption{nADI images of HD163296: same as Fig.4, but in this case we display a FOV of 0\farcs58$\times$0\farcs58, to show the faint detection at $\sim$0\farcs17 SW from the central star (magenta circle).   
}\label{hd1632_nadi_comp}
\end{figure}

\subsection{HD\,163296}\label{hd1632}

HD163296 was observed twice: in August and September 2018. In the first dataset, the observing window was not exactly within the specified  local sidereal time interval, which is why the target was reobserved later. We note that this is a difficult target for service observations from Paranal Observatory, as it crosses the meridian close to the zenith:  the target declination, -21$^{\degree}$57'21", is close to the observatory latitude, 24$^{\degree}$37'39" S. Hence, we can only obtain data before and/or after the meridian crossing, when the rotation rate is smaller. As a consequence, both datasets show a small rotation field of $\sim$7 and 9 degrees, respectively. 

In the first dataset (Aug 2018), the first 130 images were obtained with an individual exposure time (DIT) of 30 seconds, and a large fraction of them were close to (or at) the saturation level, and so we rejected them. The loop then opened, and once it was recovered,  a second sequence of 220 unsaturated images with DIT = 20 sec was obtained, but covering only  a rotation field of 1.6\,degrees.
In the second dataset (Sep 2018), we obtained a total of 255 images, each of 20 seconds exposure. In this case, the loop was more stable, and we rejected 12 images out of 255 with the lowest fluxes. We performed our analysis on the second dataset, and the final images are displayed in Figure~\ref{imagesCADI}. 
For completeness, we include an image with a larger field-of-view (FOV)
in the Appendix (Figure~\ref{Fig_HD163296_LF}), displaying the position of the planet candidates reported for this source.
Without considering the planet candidate proposed by \citet{Pinte2018} at 2.4\,arcsec (which is outside the ZIMPOL FOV), we do not detect any clear signal from the planet candidates reported in the literature. On the other hand, we detect a faint emission in the N\_Ha and ASDI images at a separation of $\sim$171\,mas SW of the central source (magenta circle, Fig.~\ref{imagesCADI}). 
Interestingly, this emission is not observed in the continuum image. 

However, we note that a small field rotation can imply important self-subtraction. We therefore also analyzed the derotated images without subtracting the PSF (named ``nADI'' here). The nADI images are displayed in Figure~\ref{hd1632_nadi} (FOV of 0.29 $\times$ 0\farcs29) and in Fig.~\ref{hd1632_nadi_comp} (FOV of 0.58 $\times$ 0\farcs58).
In the former, we can clearly see a bright point-source emission at a separation of $\sim$ 43\,mas (4.3\,au) in the SE direction (top left panel, red circle). To further investigate the nature of this emission, we also analyzed the nADI unsaturated data obtained in
2018 August 18, and checked that the bright spot appears approximately at the same location. 
Figure ~\ref{hd1632_nadi} shows that the nADI images in the individual filters show an elongated  PSF, which is distorted in the two epochs but in different directions (Fig.~\ref{hd1632_nadi}). Nevertheless, the ASDI images are very similar, displaying an elongated negative pattern in the central regions. For this reason, we think the detected bright spot is probably an artifact connected to the flux variation of a bright Airy ring with the suboptimal AO correction. 

Regarding the faint emission reported at 171\,mas, it is clearly detected in the nADI images of the two epochs ( Fig.~\ref{hd1632_nadi_comp}). As in the case of the CADI images, it is detected in the ASDI and N\_Ha images, but not in the continuum, as expected for an accreting protoplanet. 
The emission is located at 171\,mas and PA$\sim$239 degrees.
We note that this emission is not reported in \citet{Xie2020}, given that the MUSE data show strong residuals at separations $<$0\farcs5. According to \citet{Isella2018}, the possible candidate would be placed after the bright ring at 14\,au (named B14). The nature of this source (real detection or speckle) is uncertain, and would require further observations with a better rotation field coverage. Finally, we note that despite the small field rotation, the cADI is achieving very similar performances to nADI down to 50-80\,mas.

\section{Results}\label{results}

\subsection{H$_\alpha$ emission from potential protoplanet companions}

\begin{figure*}[t!]
 \sidecaption
\includegraphics[width=12cm]{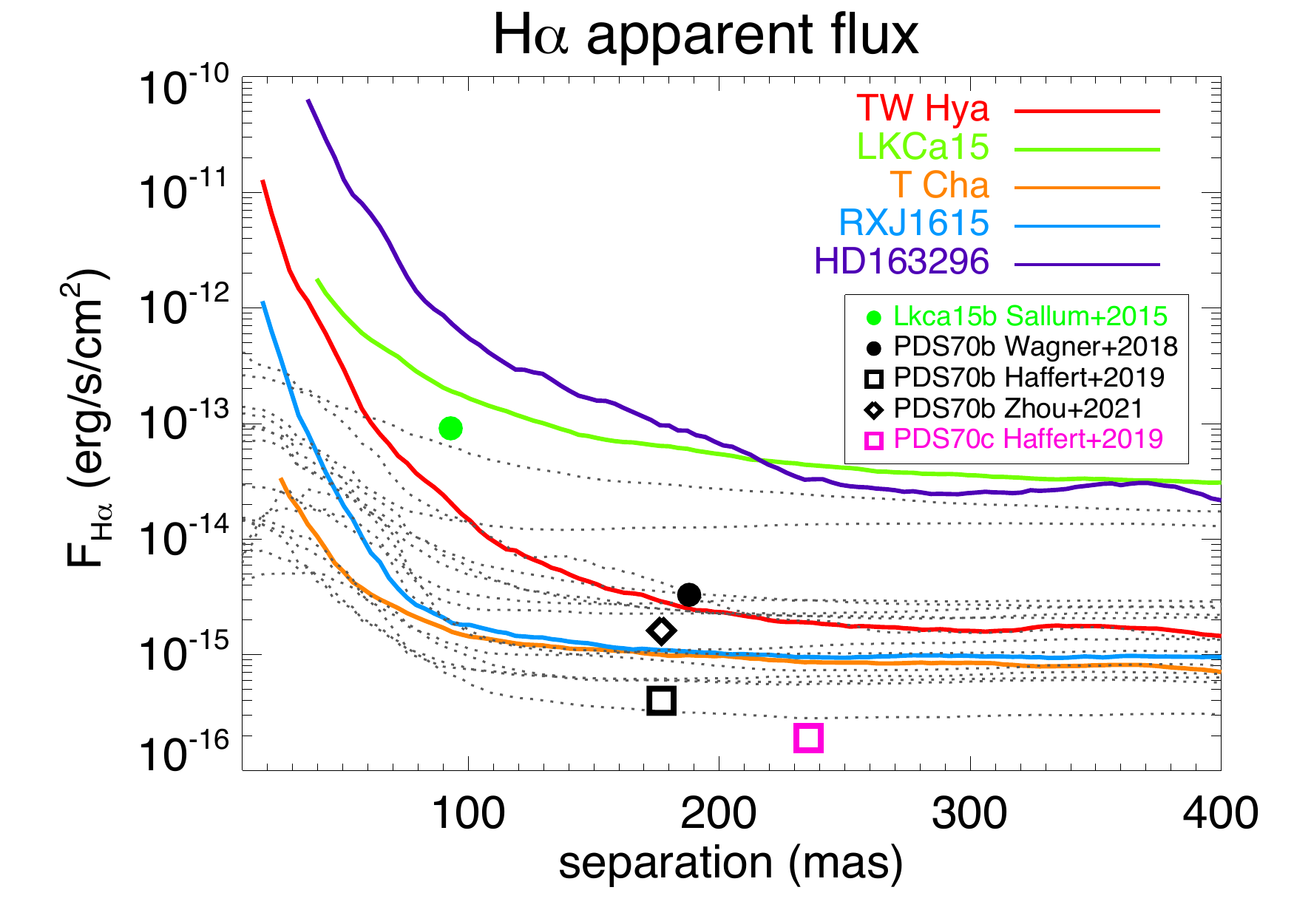}
 \caption{Apparent H$_{\alpha}$ line fluxes for the five observed targets as a function of distance, in mas, from the central star. Gray dotted lines show the fluxes of the sample observed by \citet{Zurlo2020} also with SPHERE/ZIMPOL.}\label{aflux}
 \end{figure*}

  \begin{figure*}[]
 \sidecaption
\includegraphics[width=12cm]{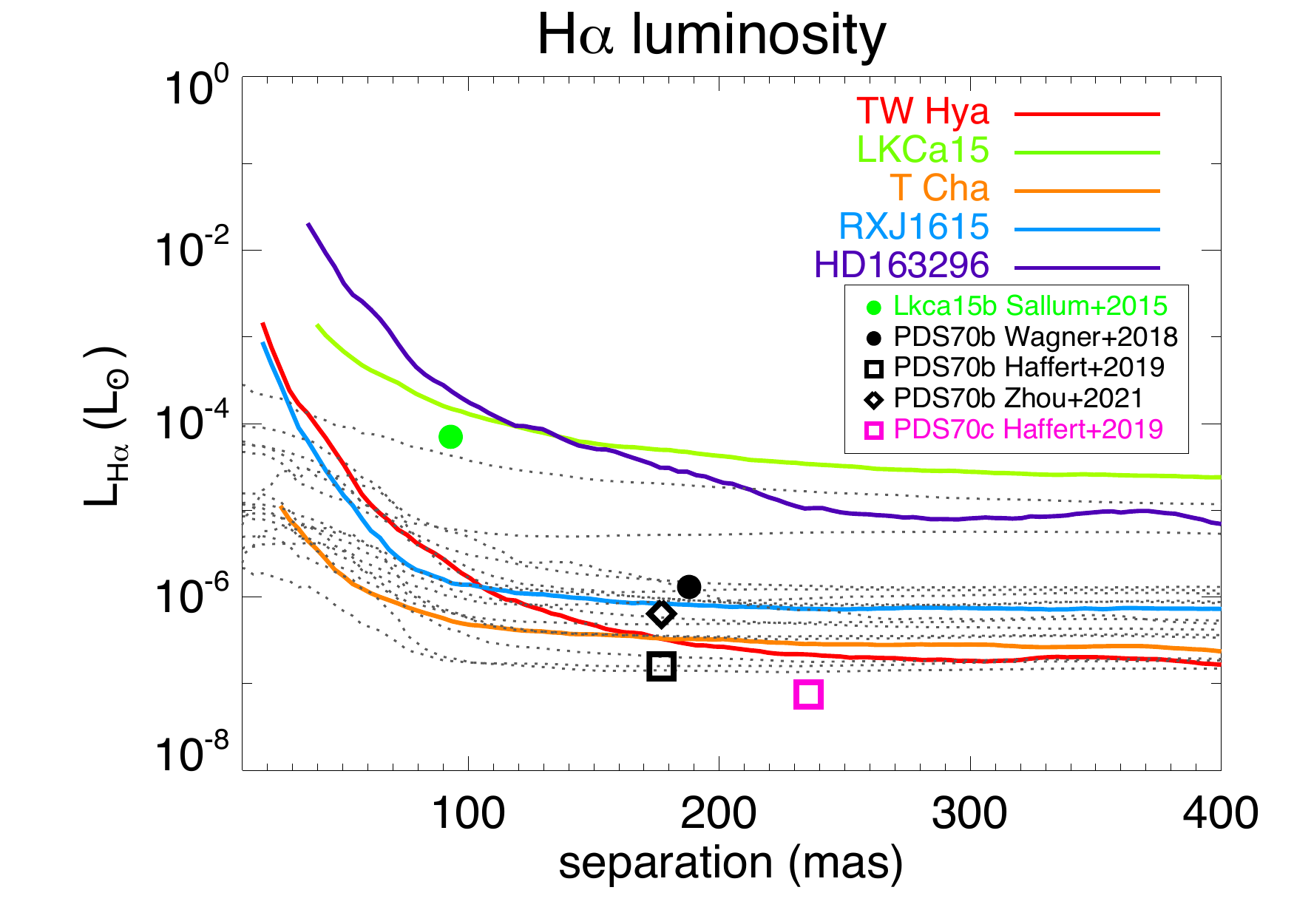}
 \caption{H$_{\alpha}$ line luminosities for the five observed targets. The gray dotted lines represent the H$_{\alpha}$ line luminosities from the sample studied by \citet{Zurlo2020}}\label{lumiALL}
 \end{figure*}
 
The high sensitivity of the ZIMPOL images  allowed us to derive N\_Ha contrast curves with $\Delta\, mag$ between of  $\sim$6 and 8 at $\sim$100\,mas from the central sources (except in the case of LkCa15). 
We can use these curves to derive upper limits to the H$_{\alpha}$ line emission of potential protoplanet companions. To this aim, we first estimated the pure H$_{\alpha}$ line flux (continuum subtracted) of the primary stars, following the methodology explained in \citet{Cugno2019}. We  first performed aperture photometry for the primaries in the ContHa and N\_Ha filters using an aperture radius of 1.5\,arcsec in all the images of the stacked cube, deriving the mean count rate ($cts$), and its uncertainty, $\sigma$/$\sqrt{n}$, with  $\sigma$ being the standard deviation, and $n$ the number of images.  We then estimated the flux in the Cnt\_Ha and N\_Ha filters following the expression:

\begin{equation}
    F_{filter} = cts_{filter} \times 10^{0.4(airm*k_{\lambda}+m_{mode})}\times c_{zp}^{filter},
\end{equation}

where $cts$ is the measured count rate (cts/s), $airm$ the average airmass (see Table~\ref{obslog}), $k_{\lambda}$  the atmospheric extinction at Paranal \citep[0.085 and 0.082 for Cnt\_{Ha} and N\_Ha, respectively,][]{Patat2011},  $m_{mode}$ (with a value of --0.23 mag) a factor taking into account the enhanced transmission of the used R-band dichroic with respect to the gray beam splitter, and $c_{ZP}^{filter}$ is the zero point of the corresponding filter. As we have observed with the Cnt\_Ha in the FW1 and N\_Ha in the FW2, the corresponding zero points are 1.59$\pm$0.05 $\times$ 10$^{-17}$ erg/(cm$^2$ $\cdot \AA \cdot$ count) and 9.2$^{+4}_{-0.5}$ $\times$ 10$^{-16}$ erg/(cm$^2$ $\cdot$ count), respectively \citep{Schmid2017}. 
The estimated flux uncertainties account for the errors in the count rates and the zero points.

Once the Cnt\_Ha and N\_Ha fluxes are estimated, we can derive the count rate due to the continuum emission in the line filter (N\_Ha), following Eq. (2) in \citet{Cugno2019}. If we subtract the derived count rate from the one measured in the line filter, we  can estimate the pure H$_{\alpha}$ emission in the N\_Ha filter for each central star (named $F_{H_\alpha}$). The derived fluxes in the two ZIMPOL filters are provided in Table~\ref{tableFlux} together with the pure H$_{\alpha}$ stellar emission in N\_Ha. 

We corrected the stellar H$_{\alpha}$ line fluxes ($F_{H_{\alpha}}$) for interstellar extinction estimating $A_{R}$ from $A_{V}$ (included in Table~\ref{stars1}), applying the extinction law from \citet{Rieke1985}.  We then used the dereddened fluxes to estimate the line luminosities ($L_{H\alpha}$) of the stars using the distance included in Table~\ref{stars1}. For completeness, we also derived the accretion properties of the central stars following the procedure described in Appendix \ref{ap_accretion}, where we also briefly discuss the obtained values.

\begin{table*}[t!]
\caption{Continuum and H$_{\alpha}$ line fluxes estimated for the primary stars and derived accretion parameters}\label{tableFlux}
\begin{tabular}{lllllll}
\hline
Target & $F_{\rm Cnt\_Ha}$               & $F_{\rm N\_Ha}$         & $F_{\rm H\alpha}$ & $L_{\rm H\alpha}$$^*$ & $L_{\rm acc}$              & $M_{\rm acc}$ \\  
           & (10$^{-12}$ erg/s/cm$^2$)   & ( 10$^{-12}$ erg/s/cm$^2$)           & ( 10$^{-12}$ erg/s/cm$^2$)  & (10$^{-3}$ L$_{\odot}$)         & (L$_{\odot}$) & (M$_{\rm \odot}$/yr) \\
\hline
RXJ1615     &   2.70$\pm$0.09 & 2.1$\pm$0.9   & 1.5$\pm$0.7   & 1.7$\pm$0.5  &  0.042 &  5.5$\times$10$^{-9}$\\
LkCa\,15    &   1.16$\pm$0.04 & 1.8$\pm$0.8   & 1.5$\pm$0.7   & 1.8$\pm$0.5  &  0.043 & 2.1$\times$10$^{-9}$\\ 
TW~Hya      &    5.4$\pm$0.2  & 29.9$\pm$13.0 & 28.7$\pm$12.5     & 3.3$\pm$1.4  &  0.085 & 5.1$\times$10$^{-9}$\\
T~Cha       &    8.3$\pm$0.2  & 2.1$\pm$0.9   & 0.3$\pm$0.1   & 0.23$\pm$0.03  &  0.004 & 2.0$\times$10$^{-10}$\\ 
HD\,163296 &     155$\pm$4    & 116$\pm$50    &  82$\pm$36    & 26.3$\pm$11.5 &  3.235 & 9.8$\times$10$^{-8}$\\  \hline 
\end{tabular}

$^*$ Dereddened values
\end{table*}

We used the dereddened H$_{\alpha}$ fluxes together with the contrast curves in the N\_Ha filter  to derive the $H_{\alpha}$ line fluxes and luminosities at different separations from the stars. We note that we used the contrast curves obtained from the cADI reduction for all the targets. 
In the case of TW~Hya, T~Cha, and RXJ1615, this could be considered a conservative approach for the innermost regions, but this is not critical.
The flux detection limits and luminosity upper limits are shown in Figures~\ref{aflux} and~\ref{lumiALL}, respectively. 
For comparison, we include the curves obtained for the sample of objects observed by \citet{Zurlo2020} with SPHERE/ZIMPOL.
As shown in  Fig.~\ref{lumiALL}, for three objects (T Cha, TW Hya, and RXJ1615), we obtained H$_{\alpha}$ line luminosities of  $\sim$0.5--1$\times10^{-6}\, L_{\odot}$ at separations of 200\,mas, which is comparable to the values derived by \citet{Zurlo2020} for most of their sources. In the case of HD\,163296 and LkCa15, the estimated line luminosities are $\sim$3$\times10^{-5}\,L_{\odot}$ and $\sim$1$\times10^{-4}\,L_{\odot}$ at 200\,mas, respectively. Also, for the sake of comparison, we include the H$_{\alpha}$ fluxes of PDS70\,b and c, and LkCa15\,b in Fig.~\ref{aflux}.

For each target, we estimated contrasts, H$_{\alpha}$ line fluxes, and  luminosities at the projected separations where the innermost dust gaps (corresponding to the disk semi-major axis) or planet candidates have been detected (see Table~\ref{planets}). We note that these are rough estimations, because we have not corrected the curves by the disk inclinations. The results are included in Table \ref{planets}. For the dust disk gaps (marked with a ``{\em g'}'), and whenever available, we have included the planet mass that could be responsible to carve it. 
For the planet candidates detected in IR imaging or submm ALMA data (marked with a ``{\em p'}'), we have included the planet mass estimated from the corresponding observations.
As seen in Table~\ref{planets}, 
for the targets with good-quality data, we reach upper limits to the H$_{\alpha}$ line luminosity of 10$^{-6}$\,L$_{\odot}$ in the innermost gaps, and $\sim$10$^{-7}$\,L$_{\odot}$ in the outermost ones. For LkCa\,15 and HD\,163296, these upper limits are between 10$^{-4}$\,L$_{\odot}$ and 10$^{-6}$\,L$_{\odot}$, respectively.

\begin{table*}
\caption{5-$\sigma$ upper limits to the N\_Ha contrast, dereddened H$_{\alpha}$ line fluxes and luminosities, and accretion luminosities at the projected separations of dust gaps (corresponding to the disk semi-major axis), and planet candidate detections.}\label{planets}
\begin{tabular}{lllcllll}
\hline
Target & \multicolumn{2}{c}{Projected Separation}  & Planet mass & N\_Ha contrast  & $F_{\rm H\alpha}$ & $L_{\rm H\alpha}$ & $log L_{\rm acc}$$^{\dagger}$ \\  
           & (arcsec)   & (au)$^{*}$   & (M$_{\rm Jup}$) & (mag) & (10$^{-15}$ erg/s/cm$^2$)  & (10$^{-6}$ L$_{\odot}$) & (L$_{\odot}$) \\
\hline
RXJ1615     & 0.14$^{1, (g)}$     & 22 & 4.5$^1$ & 8.1 & $<$ 1.3  &$<$ 0.98 & $<$ -4.1\\
            & 0.50$^{2, (g)}$    & 78 & 0.2$^{3}$ & 8.6 & $<$ 0.8 &$<$ 0.64 & $<$ -4.3\\ \hline
LkCa\,15       &  0.093$^{4, (p)}$    & 15  & 6$^{5}$ &  2.7 & $<$ 190 & $<$ 150  & $<$-2.0\\   
               & 0.27$^{6, (g)}$ & 43 & 0.5$^{3}$ & 4.4 & $<$ 38 & $<$ 0.29 & $<$ -2.7 \\ \hline
TW Hya         &  0.10$^{7, (g)}$     &  6  & 0.04$^{1}$ & 8.2 & $<$ 14  & $<$ 1.6 & $<$ -3.9\\
               &  0.37$^{7, (g)}$     &  22 & 0.15$^{3}$ & 10.6 & $<$ 1.7 & $<$ 0.19 & $<$ -4.8\\ 
               &  0.87$^{8, (p)}$     &  52 & 0.05$^8$ &   11.1 & $<$ 1.0 &  $<$0.11 & $<$ -5.0     \\ \hline
T~Cha       &  0.093$^{9, (g)}$ & 9    & 1.2$^9$  & 6.0 & $<$ 1.6 & $<$ 0.52   & $<$ -4.4 \\ \hline
HD\,163296  &  0.10$^{10, (g)}$     & 10 &  &  5.4  & $<$ 700 & $<$ 220  & $<$ -1.9 \\ 
            &  0.44$^{11, (g)}$  & 45 &  &   9.5  & $<$ 16 & $<$ 5.2 & $<$ -3.4 \\
            &  0.5$^{12, (p)}$   & 51 & 6$^{12}$ & 9.9 & $<$ 10 & $<$ 3.3 & $<$ -3.6 \\
            &  0.67$^{13, (p)}$  & 68 & 1$^{13}$ & 10.6 & $<$ 5.6 & $<$ 1.8 &  $<$ -3.8 \\
            & 0.77$^{14, (p)} $  & 78 & 1$^{14}$  & 10.8  & $<$ 4.5 & $<$ 1.4 & $<$ -3.9\\      
              \hline
\hline
\end{tabular}

$^{*}$ Estimated assuming Gaia EDR3 distances from Table~\ref{stars1}; 
$^{\dagger}$ Estimated using the prescription of \citet{Aoyama2021}  (see text).
$^{(g)}$: Disk gaps; $^{(p)}$: Planet candidates;

$^{1}$ \citet{ATorres2021}; 
$^2$ \citet{deBoer2016};
$^{3}$ \citet{Dong2017}, values for a viscosity parameter $\alpha$ 10$^{-3}$;  
$^4$ \citet{Sallum2015};
$^5$ \citet{Kraus2012};
$^6$ \citet{Thalmann2016};
$^7$ \citet{vanBoekel2017}; 
$^8$ \citet{Tsu2019}; 
$^9$ \citet{Hendler2018}, position of the gap minimum; 
$^{10}$\citet{Isella2018}; 
$^{11}$ \citet{Isella2016}; 
$^{12}$ \citet{Guidi2018};
$^{13}$ \citet{Pinte2020}; 
$^{14}$ \citet{Izquierdo2021}

\end{table*}

\subsection{Accretion luminosity of potential planet companions}\label{planetsACC}

The accretion luminosity ($L_{\rm acc}$) of young planets is an important parameter as it be used  to estimate their accretion rate, and therefore to understand how planets grow as they evolve. The $L_{\rm acc}$ of planet candidates can be estimated from the observed  $H_{\alpha}$ line emission, after assuming a model to explain the physical conditions to  form the line. To illustrate the impact of the different models on the $L_{\rm acc}$ estimations, we compared three different approaches, and applied them to the sources observed with SPHERE/ZIMPOL.

Previous works \citep[e.g.,][]{Close2014,Sallum2015, Cugno2019} calculated the accretion luminosity of young giant planets assuming that it scales with $L_{\rm H_{\alpha}}$ as in low-mass Classical T Tauri stars (CTTSs). They therefore applied the relation derived by \citet{Rigliaco2012}: 

\begin{equation}\label{laccTTS}
 \log (L_{\rm acc}/L_{\odot}) = (2.99\pm0.16) + (1.49\pm0.05)\times \log(L_{\rm H\alpha}/L_{\odot}). 
\end{equation}

However, as already mentioned in different works, this relation is not necessarily valid for young planets, because CTTSs show different properties \citep[see e.g.,][]{Zhou2014,Thana2019}, and the accretion process can take place through a different mechanism.  
We include this approach here so that our results can be compared  with previous works, but it is important to estimate $L_{\rm acc}$ using  appropriate models developed for young planets, and not for young stars. 

 In this context, \citet{Aoyama2018} developed a 1D model of shock-heated gas for planetary masses, and studied the hydrogen line emission from young giant planets. In their model, the $H_{\alpha}$ line emission is generated in the post-shock gas located either in the CPD or on the planet surface \citep{Aoyama2019}.  In an extension of this work, and using a broad range of planetary masses and accretion rates, \citet{Aoyama2021} provided a $L_{H_\alpha} - L_{\rm acc}$ relationship (Eq.~\ref{laccAoyama})  assuming a magnetospheric accretion model and considering that the accretion shock is located on the planetary surface:

\begin{equation}\label{laccAoyama}
 \log (L_{\rm acc}/L_{\odot}) = 1.61 + 0.95\times \log(L_{\rm H\alpha}/L_{\odot}), 
 \end{equation}
for $L_{\rm acc} \le 10^{-4} L_{\odot}$ , although, as explained by the authors, higher values barely affect the fit. The half-spread at a given $L_{H_{\alpha}}$ is of 0.3\,dex. 

A different model was presented  by \citet[][SE2020 hereafter]{Szu2020}, in which planets grow through boundary layer accretion \citep[see e.g.,][]{Owen2016} in which the 
material is accreted directly onto the planet and its CPD from the circumstellar disk.
SE2020 computed hydrogen recombination line emission from 3D thermo-hydrodynamical models of forming planets with masses of 1-10\,M$_{\rm Jup}$, incorporating extinction, line variability, and excluding magnetic fields. In their 3D simulations, 
all the line luminosity comes from the accretion shock on the surface of the CPD.
SE2020 provided parametric equations between the accretion luminosity and the H$_{\alpha}$ line emission for different size distributions and chemical compositions of the circumstellar dust grains.  For our comparison, we considered two different opacity cases included in their model: a disk with a dust mixture of silicates, water ice, and carbon (the most realistic case in terms of composition), and a gas-only disk, that is, a dust-depleted disk to consider the largest disk gaps where planets can grow, and which are almost devoid of dust. We note that the latter is an extreme scenario and will only provide upper limits to the line luminosities. SE2020 derived the following relationships for these two opacity cases:

\begin{equation}
    \log(L_{\rm acc}/L_{\odot}) = 0.028 \times \log (L_{H_{\alpha}}/L_{\odot}) -2.55 
    \end{equation}\label{eqSzu1}
for the disk with a mixture of silicates, water ice, and carbon, and 

\begin{equation}\label{eqSzu2}
 \log(L_{\rm acc}/L_{\odot}) = 0.32 \times \log (L_{H_{\alpha}}/L_{\odot}) -2.18  
\end{equation}
for the gas-only case.

\begin{figure*}[t!]
  \sidecaption
\includegraphics[width=12cm]{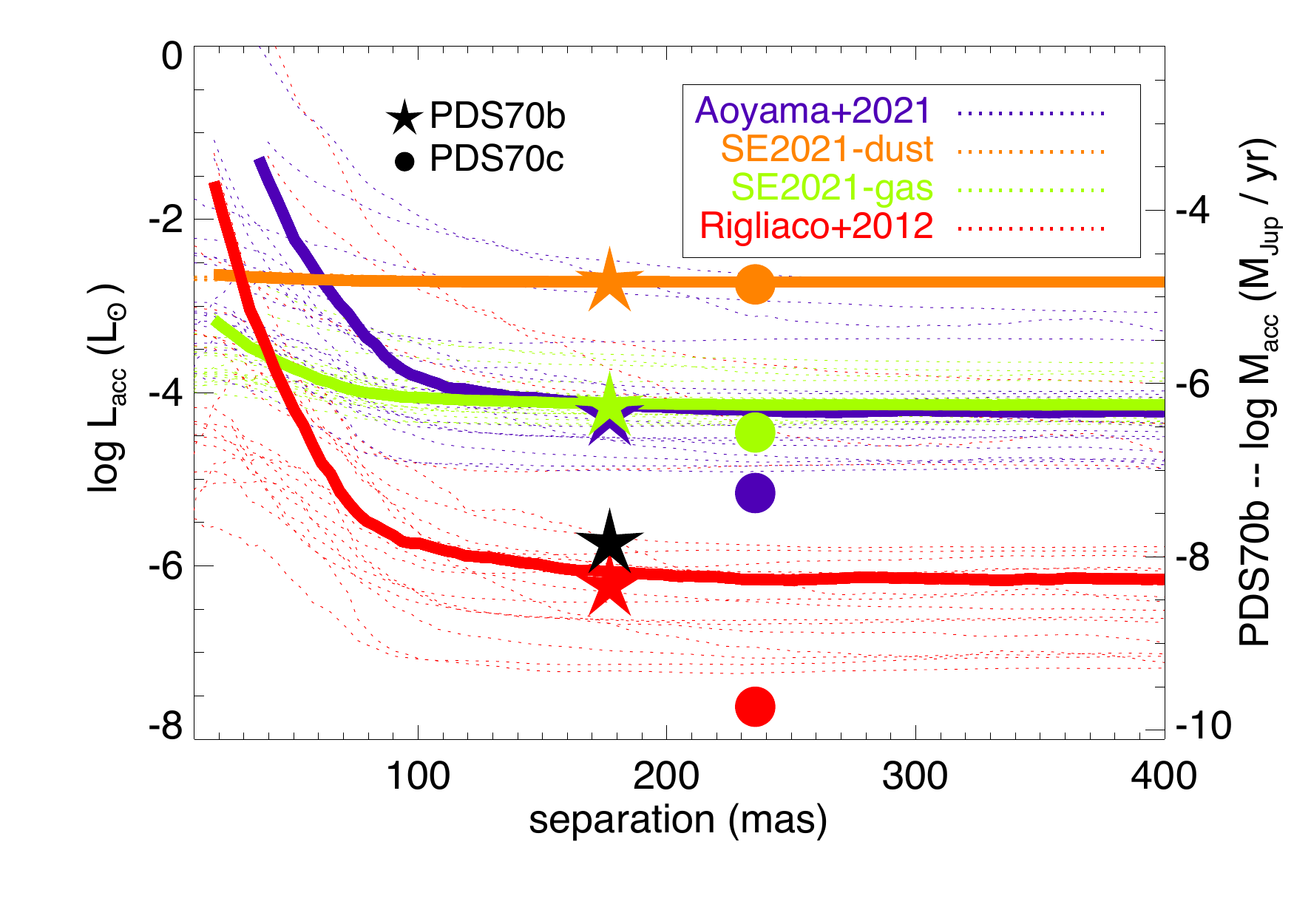}
\caption{Accretion luminosities estimated from the $L_{\rm H_{\alpha}}$ contrast curves, and using four different relations given in Eqs.~\ref{laccTTS} -- \ref{eqSzu2} (see text). These four relationships are represented with different colors as described in the upper right panel of the figure.  The dotted lines correspond to all the individual objects observed in this work and in \citet{Zurlo2020}, while the thick solid lines represent the results for RXJ1615 (as an average case) to highlight the difference between the four models. We represent the estimated $L_{\rm acc}$ values for planets PDS70b (starred symbols) and PDS70c (filled circles) using the four models. For completeness, we include the  $L_{\rm acc}$ value of PDS70b estimated by \citet{Zhou2021} using a slab model (black star). The corresponding $M_{\rm acc}$ values for planet PDS70b, assuming a planet mass of 1\,M$_{\rm Jup}$ and a planet radius of 1.75\,R$_{\rm Jup}$, are displayed on the right y-axis.}\label{Lacc_lha}
 \end{figure*}

In order to translate our $L_{H_{\alpha}}$ observational limits into accretion luminosities, we estimated $L_{\rm acc}$ considering all the models described above, and the 5-$\sigma$ $H_{\alpha}$ luminosity curves of the objects included in this work and in \citet{Zurlo2020}. The results are represented in Figure~\ref{Lacc_lha}, where we  highlight the results obtained for RXJ1615 (average case) to see the differences between the models more clearly. 

 As seen in Fig.~\ref{Lacc_lha}, for a given $L_{H_{\alpha}}$, the planetary models predict accretion luminosities higher than those previously calculated using stellar relations. If we considered a separation of for example 200\,mas, the estimated accretion luminosity is on average two orders of magnitude higher (or almost three times in the case of the dusty disk from SE2020) than the $L_{\rm acc}$ derived with the relation of  \citet{Rigliaco2012}. Therefore, when using planetary models, an observed $L_{H_{\alpha}}$ would correspond to a higher $M_{\rm acc}$  than previously estimated for a given planet mass and radius. As explained by \citet{Aoyama2021}, this could indicate that only the strongest accretors can produce an H$_{\alpha}$ emission line bright enough to be detected, and could explain the lack of detections of the different surveys.
 
To illustrate the impact of using different models, we estimated the $L_{\rm acc}$ of the confirmed protoplanets PDS70bc using the four approximations described above. In the case of PDS70b, we consider a line luminosity of $L_{H_{\alpha}} = 6.5\times 10^{-7}\,L_{\odot}$ \citep{Zhou2021}, while in the case of PDS70c we use the line flux obtained by \citet{Haffert2019}. The results are included in Fig.~\ref{Lacc_lha}.
For completeness, we also include the $L_{\rm acc}$ value of PDS70\,b derived by \citet{Zhou2021}, who fit UV and H$_{\alpha}$ observations with a slab model. Clearly, the different models can provide $L_{\rm acc}$ that can differ by more than three orders of magnitude, which 
translates to a very different accretion rate. For example, in the case of PDS70b, considering a  mass of 1\,M$_{\rm Jup}$ and a radius of 1.75\,R$_{\rm Jup}$ \citep{Stolker2020}, we can estimate the accretion rate following \citet{Gullbring1998}, and assuming $R_{\rm in}$ = 5\,$R_p$ as in \citep{Zhou2021}. The derived values are displayed on the right y-axis of Figure~\ref{Lacc_lha}. This graph illustrates that, depending on the model used, the derived accretion rates range between $\sim$ 1$\times$10$^{-8}$ and 1$\times$10$^{-5}$ M$_{\rm Jup}$/yr, meaning that the protoplanet can be classified as  either a strong or faint accretor.

Finally, and for completeness, 
we estimated the $L_{\rm acc}$ from the $L_{H_{\alpha}}$ for the planet candidates and gaps detected around the five sources studied in this work. For this estimation, we used the prescription of \citet{Aoyama2021} (Eq.~\ref{laccAoyama}). The values are included in the last column of Table~\ref{planets}, and range between $\log(L_{\rm acc})\sim-2.0$ and $-5.0\, L_{\odot}$ for the different sources and separations.

\section{Discussion}

\citet{Cugno2019} and \citet{Zurlo2020}  carried out ZIMPOL/H$_{\alpha}$ surveys to detect accreting protoplanets in the gaps of transitional disks. Adding up the new targets observed in this study, there are a total of 18 objects observed with ZIMPOL:
8 T Tauri stars and 10 Herbig Ae/Be and F-type stars. Although most of these ZIMPOL observations were obtained under good conditions, reaching a very high contrast, no accreting protoplanets were detected. The only companion detected with ZIMPOL was around the source HD\,142527, confirming its previous H$_{\alpha}$ detection \citep{Close2014}.

Other works using different instrumentation have reported similar results: \citet{Xie2020} did not detect any new accreting protoplanet around five young stars using MUSE on the VLT, even though this instrument can reach fainter detection limits in the H$_{\alpha}$ line apparent flux (a factor of $\sim$5 fainter) in comparison to SPHERE/ZIMPOL.  Finally, \citet{Uyama2020} observed five T Tauri stars with Subaru/SCExAO+VAMPIRES, reporting the nondetection of nearby planet companions.

We should note that most of these surveys reach the highest contrast at separations of larger than $\sim$ 100-150\,mas, which translates to projected separations of larger than 10-15\,au for nearby objects. The deprojected separations can be even larger, which means that we obtained the best sensitivity in regions where giant planets are difficult to form. On the other hand, the only protoplanets confirmed so far are detected at large projected separations from the central star: 0\farcs195 (22\,au) and 0\farcs22 (30\,au) 
for PDS70\,b and c, respectively \citep{Keppler2018, Mesa2019b}. 
Despite the fact that the ZIMPOL observations from different surveys provide very good contrast at similar separations for some of the observed targets, no protoplanets were
detected in the same spatial range.

Apart from the lack of sensitivity at very small separations,
 the lack of H$_{\alpha}$ detections could be explained by a combination of other factors; for example, weak and/or episodic accretion, circumstellar and circumplanetary disk extinction, and/or the presence of planets with insufficient mass to produce a detectable H$_{\alpha}$ line emission. 

As explained in the previous section, a large number of works estimating accretion luminosities and accretion rates in protoplanets used relations derived for TTSs, which are not valid in the planetary regime. When using planetary models, the result is that only the strongest accretors could be detected in the $H_{\alpha}$ line, and this could explain why previous surveys reported a large number of nondetections. 
We compared this result with the predictions from planet population synthesis performed by \citet{Mordasini2017}.
In the ``cold accretion'' scenario described by these latter authors, the accretion luminosity caused by shocks is radiated away from the planet, meaning that they can derive the accretion luminosities as a function of planet mass for the planet population at 3\,Myr (their Figure~7, cold-nominal case). According to the results from planetary models, 
current H$_{\alpha}$ observations would only be sensitive to either planets more massive than  1\,M$_{\rm Jup}$ with accretion luminosities higher than 10$^{-4}$\,$L_{\odot}$,  
or to the most massive planets with M$_{p}$ $>$ 10\,M$_{\rm Jup}$ (a very reduced sample),  assuming the dust model (SE2021-dust) by \citet{Szu2020}.
In this context, PDS\,70 b might fit with the scenario described by \citet{Mordasini2017} as a limiting case,  given its mass (1.0$\pm$0.5\,$M_{\rm Jup}$) and the estimated accretion luminosity when planetary models are considered (see Figure~\ref{Lacc_lha}).  In the case of planet PDS70\,c, with an estimated mass of lower than 5\,M$_{\rm Jup}$ \citep{Mesa2019b}, it could fit in this scenario only if $L_{\rm acc}$ is estimated using the models by \citet{Szu2020}.

Regarding the episodic accretion, there is only one study that has monitored the H$_{\alpha}$ line emission in the protoplanet PDS70\,b: \citet{Zhou2021} observed the planet in six epochs over a five-month timescale with the HST, and their high-quality data did not support a H$_{\alpha}$ line variability at a level higher than 30\%. In the same respect, it would be extremely useful to perform a dedicated campaign to study the long-term variability of the H$_{\alpha}$ line emission of the two confirmed protoplanets PDS70\,b and c.

Another parameter that can affect the measured $H_{\alpha}$ luminosity  is the mass of the accreting planets. As explained in SE2020, there is a trend between planet mass and line luminosity, as the temperature is higher in the vicinity of a more massive planet (resulting in a higher hydrogen ionization), and the extinction decreases because the gap opened is larger (see below). As a result, these models predict that when considering the most realistic disk opacities, only giant planets with masses above 10\,M$_{\rm Jup}$  show a clearly detectable H$_{\alpha}$ emission line at the level of $L_{H_{\alpha}} \sim 10^{-2} L_{\odot}$. When taking into account line variability due to extinction and/or episodic accretion, SE2020
show that 5\,M$_{\rm Jup}$ planets can be detected with $L_{H_{\alpha}} \sim 10^{-8}-10^{-5} L_{\odot}$, some of these values being within the ZIMPOL detection limits. 
We note that in the present work, we did not consider the extinction due to the dust in the circumstellar disks and/or the  CPD when estimating $L_{H_{\alpha}}$. 

In the case of transitional disks, \citet{Sanchis2020} studied the effect of circumstellar extinction on giant planets, and concluded that it varies depending on the mass of the object because this parameter defines the width and depth of the opened gaps. According to their estimations, planets with masses of 1-2 $M_{\rm Jup}$ would show very high extinction, while planets with 5\,M$_{\rm Jup}$ or higher, would show a negligible extinction.
In our sample of five stars, the predicted planet masses range between $\sim$ 16 M$_{\rm Earth}$ and 6\,M$_{\rm Jup}$ (see Table~\ref{planets}). 
Excluding LkCa15\,b because of the controversy surrounding its existence, the most massive protoplanet candidate is that of 6\,$M_{\rm Jup}$ around HD163296 proposed by \citet{Guidi2018}  based on IR observations. We did not detect any source at the position of this candidate, although we are limited by the small rotation field (although the nADI images do not reveal any source). Another interesting source is RXJ1615, with a predicted 4.5 M$_{\rm Jup}$ planet at 22\,au according to \citet{ATorres2021}. However, and regardless of the good contrast, we do not detect any accreting source at the given separation.

Even in the case of large gaps in the circumstellar disks, we still have to consider the extinction related to the CPDs, given that the H$_{\alpha}$ flux could be more or less absorbed depending on the CPD inclination with respect to the observer. We do not know a priori the CPD inclinations, but if these are similar to those of primary disks, we should have been able to detect some protoplanets, because a large fraction of the stars observed in the different surveys are surrounded by disks with both low and intermediate disk inclinations. The protoplanets PDS70\,bc, which are surrounded by CPDs \citep[see][]{Chris2019,Benisty2021}, were detected inside the gap of a disk with an inclination of $\sim$52 degrees \citep{Keppler2018}. In this context,  the case of T Cha is the most challenging given that the disk inclination is 67 degrees.
We can therefore conclude that the combination of some/all of the factors explained above could explain the lack of detections of accreting planets around the explored objects.

 Finally, we note that all of the surveys and works focusing on the search of accreting protoplanets through H$_{\alpha}$ observations have studied stars in the Class~II stage. However, there is already observational evidence of disks around Class~I protostars displaying similar substructures (e.g., rings, gaps, cavities) to those observed in Class~II objects \citep[e.g.,][]{Segura2020, Sheehan2020}. In addition, evidence of grain growth, a key element to planet formation, has been suggested in at least one Class I object \citep{Harsono2018}. 
Another possible scenario to explain the lack of H$_{\alpha}$ detections would therefore be that planet formation takes place earlier than predicted by models, meaning that the bulk of the accretion luminosity during the planet formation process occurs sooner in their evolution when stars are still embedded in their parental clouds.  However, further investigations should clarify whether Class~I disk substructures are indeed caused by planets or are the result of other processes (e.g., magneto-rotational instabilities, snowlines). Nevertheless, any scenario would still have to explain the detection of protoplanets PDS70\,b and c around a relatively old star.

\section{Conclusions}

We present high-angular-resolution $H_{\alpha}$ observations of five disks with signatures of planet formation. As found in previous works using SPHERE/ZIMPOL, we do not report any clear detection of accreting planets regardless of the good instrumental performance. We only report a very faint H$_{\alpha}$ point-like emission around TW~Hya, a probable speckle related with a spike, and HD163296 based on  emission detected in the ADI N\_Ha and ASDI images, but additional observations are required to confirm its true nature. The lack of detections could be explained by different factors such as low contrast at smaller separations ($<$150\,mas) where giant planets are more likely to form, a majority of low-mass, low-accreting planets, circumplanetary extinction, and/or episodic accretion.

The accretion luminosities derived from the upper limits to the H$_{\alpha}$ luminosities vary significantly when using relations from different planetary models and, in general, they are all higher than the ones derived using stellar relations from CTTs.

Future observations with other facilities such as the James Webb Space Telescope, or with ground-based telescopes with more advanced AO systems, such as MagAO+ \citep[see][]{Close2020} or SPHERE+, will help us to access more targets and to study different accretion tracers (e.g., $H_{\alpha}$, Pa $\beta$ or/and Br $\gamma$ lines), allowing us to increase the detection of accreting protoplanets in the gaps of transitional disks. Dedicated monitoring campaigns of detected protoplanets in these tracers will also help us to understand their short- and long-term variability, and to study episodic accretion in more detail.

 \begin{acknowledgements}

 This research has been partially funded by the Spanish MCIN/AEI/10.13039/501100011033 grant PID2019-107061GB-C61 and MDM-2017-0737 Unidad de Excelencia {\em Mar\'{\i}a de Maeztu} - Centro de Astrobiolog\'{\i}a (CSIC-INTA). NH is very grateful to the Paranal staff that performed the service observations, and to the ESO USD, in particular, Henri Boffin. Part of this work has been carried out within the framework of the National Centre for Competence in Research PlanetS supported by the Swiss National Science Foundation. 
 IM is funded by a RyC2019-026992-I grant by MCIN/AEI /10.13039/501100011033.
 JMA acknowledges financial support from the project PRIN-INAF 2019 {\em Spectroscopically Tracing the Disk Dispersal Evolution}. 
 GC thanks the Swiss National Science Foundation for financial support under grant number 200021\_169131.
 IdG-M acknowledges support from the Spanish MCIN/AEI/10.13039/501100011033 through grant PID2020-114461GB-I00.
 AZ acknowledges support from the FONDECYT Iniciaci\'on en investigaci\'on project number 11190837.
 This work has made use of data from the European Space Agency (ESA) mission {\it Gaia} (\url{https://www.cosmos.esa.int/gaia}), processed by the {\it Gaia} Data Processing and Analysis Consortium (DPAC,
\url{https://www.cosmos.esa.int/web/gaia/dpac/consortium}). Funding for the DPAC has been provided by national institutions, in particular the institutions
participating in the {\it Gaia} Multilateral Agreement.

\end{acknowledgements}

 \bibliographystyle{aa}
 \bibliography{zimpol}

\begin{appendix}

\listofobjects

\section{TW Hya: comparison of final images from different processing algorithms}\label{twhya_app}

\begin{figure*}[t!]
 \centering
\includegraphics[width=4.3cm,angle=270]{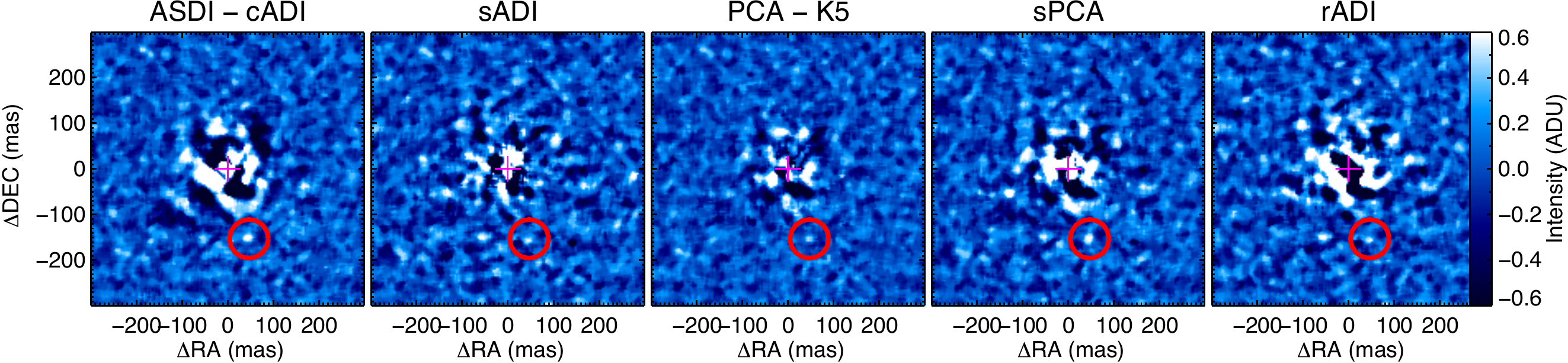}
\includegraphics[width=4.3cm,angle=270]{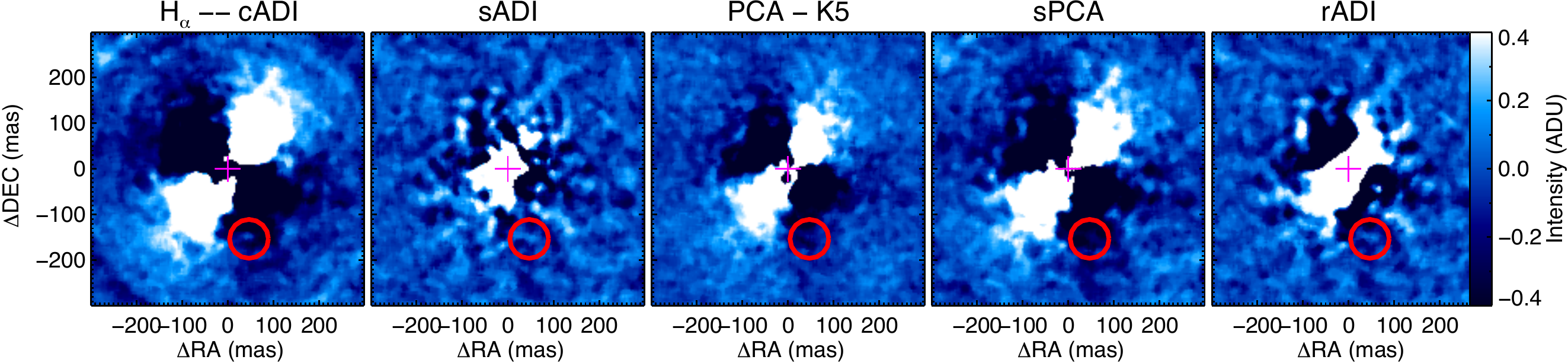}
\includegraphics[width=4.3cm,angle=270]{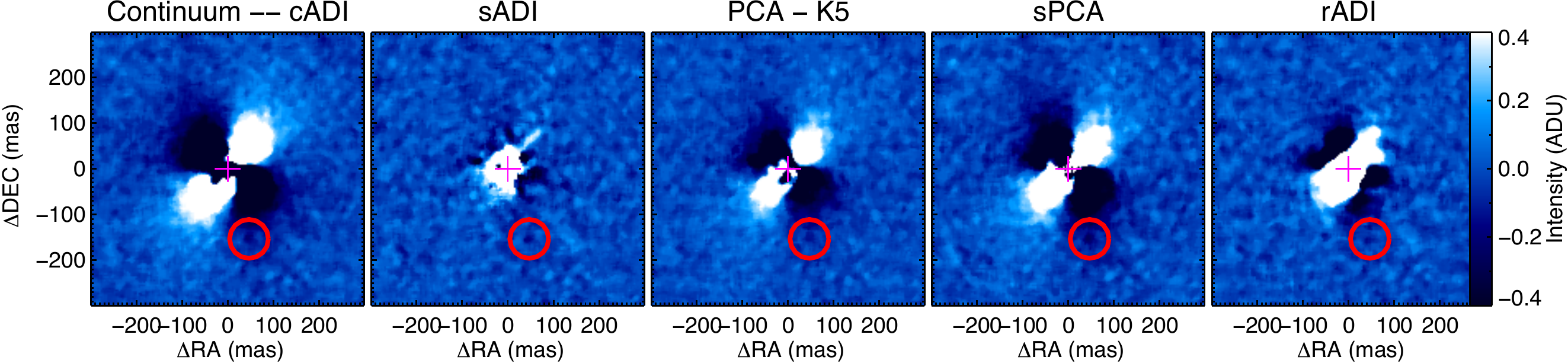}
  \caption{ASDI (top) and ADI N\_Ha (middle) and CntHa (bottom) images of TW Hya using different processing techniques. In the case of PCA processing, we show the results obtained when considering five components (denoted with K5). The bright spot detected at $\sim$160\,mas SW is encircled in all the images.}\label{twhya_all}
 \end{figure*}

This Appendix shows the result of applying the different processing algorithms explained in section \ref{analysis} to the case of the star TW Hya. This object is of particular interest because of the detection of faint point-like emission at a separation of $\sim$160\,mas (9.1\,au) from the central source. The results are displayed in Figure~\ref{twhya_all}. The upper panels correspond to the ASDI images, 
while the middle and lower panels represent the ADI N\_Ha and CntHa images.

\section{HD163296: position of the published planet candidates}

\begin{figure*}[t!]
 \centering
\includegraphics[width=16.2cm,angle=0]{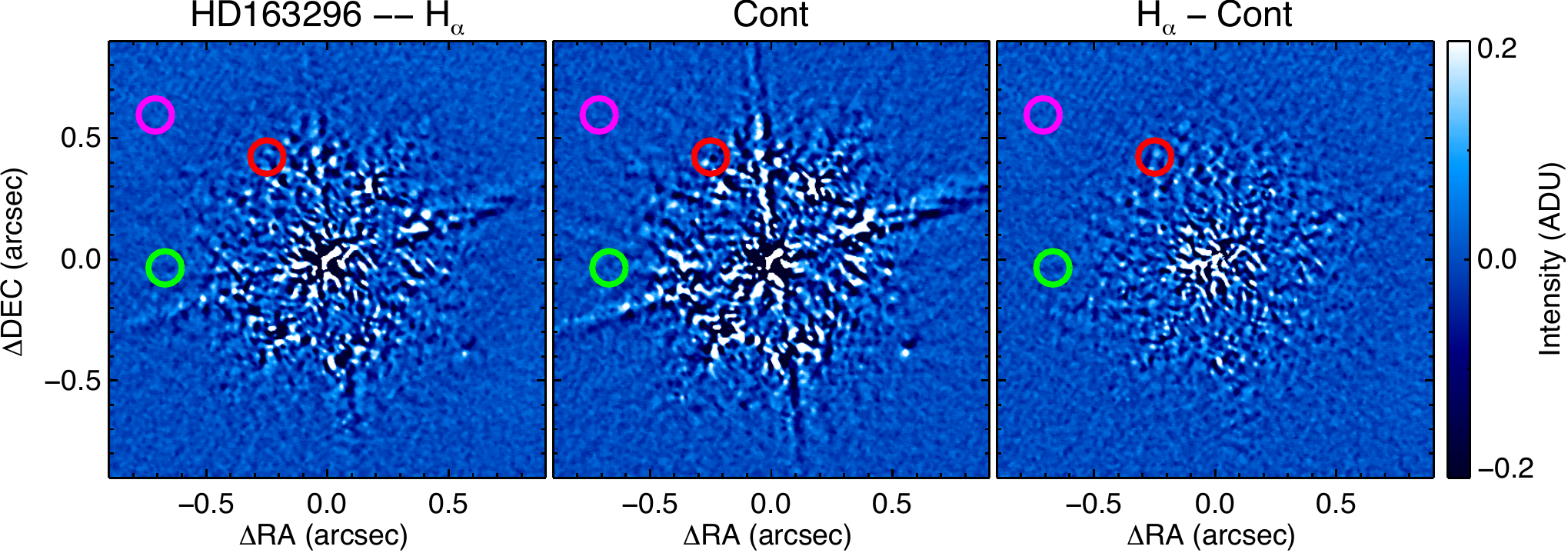}
  \caption{ADI N\_Ha (left), ADI CntHa (middle) and ASDI (right) images of HD163296 displaying a FOV of 0\farcs9 $\times$ 0\farcs9. The planet candidates reported by \citet{Guidi2018}, \citet{Pinte2020}, and \citet{Izquierdo2021} are represented by red, green, and magenta circles, respectively. We note that the planet candidate reported by \citet{Pinte2018} at 2\farcs4 is outside the ZIMPOL FOV. }\label{Fig_HD163296_LF}
\end{figure*}

 Figure~\ref{Fig_HD163296_LF} includes the ZIMPOL images of HD163296, where we represent the position of the planet candidates reported for this source in previous works.

\section{Accretion rates of the primary stars}\label{ap_accretion}

 In Section \ref{results} we estimate the H$_{\alpha}$ line fluxes of the primary stars in the sample. We compare the derived values with those previously obtained in different works and we also estimate their accretion luminosities and accretion rates. 

For the T Tauri stars in the sample, we estimated the accretion luminosity ($L_{\rm acc}$) following the relationship derived by \citet{Alcala2017}:

\begin{equation}\label{laccTTSalcala}
 \log (L_{\rm acc}/L_{\odot}) = (1.74\pm0.09) + (1.13\pm0.07)\times \log(L_{\rm H\alpha}/L_{\odot}). 
\end{equation}

We then estimated the accretion rate ($M_{\rm acc}$) of all the objects using the expression from \citet{Gullbring1998}:

\begin{equation}\label{maccTTS}
M_{\rm acc} = 1.25\times\frac{L_{\rm acc} R_*}{G M_*},
\end{equation}

where $M_*$ and $R_*$ are the stellar mass and radius (included in Table~\ref{stars1}), respectively. The expression assumes a disk truncation radius of 5\,$R_{\odot}$.

 In the case of HD163296, we estimated both $L_{\rm acc}$ and $M_{\rm acc}$ using the relations derived by \citet{Fairlamb2017} for Herbig Ae/Be stars: 

\begin{equation}\label{laccHAB}
\log(L_{\rm acc}/L_{\odot}) = (2.09\pm0.06) + (1.00\pm0.05)\times \log(L_{H\alpha}/L_{\odot}), 
\end{equation}

and

\begin{equation}\label{maccHAB}
M_{\rm acc} = \frac{L_{acc}\,R_{*}}{G\,M_{*}}
\end{equation}

We include the derived $L_{\rm acc}$ and $M_{acc}$ (expressed in $M_{\odot}$/yr) in Table~\ref{tableFlux}.  We briefly discuss the results obtained for the individual objects below:

 {\bf RXJ1615}: The estimated H$_{\alpha}$ line flux for this T Tauri star is  1.5$\pm$0.7 $\times 10^{-12}$ erg/s/cm$^2$. Using the stellar values included in Table~1, we estimated an accretion rate of 9.8$\times$10$^{-9}$\,M$_{\odot}$/yr. The line flux is consistent with the value derived by \citet{Manara2014}, 2.24$\pm$0.05 $\times 10^{-12}$ erg/s/cm$^2$, within the uncertainties.
 
 For this particular object, and to test our photometric calibration, we performed high-resolution spectroscopy of RXJ1615  with the Magellan Echellette Spectrograph (MagE)  mounted on the 6.5m Baade telescope in Las Campanas Observatory 10 days after the ZIMPOL observations. We obtained two spectra of 60 seconds each with the 5 arcsec slit on April 30 2018 ---10 days after the ZIMPOL data was obtained--- under photometric conditions. The target was observed at airmass 1.02. A spectrophotometric standard star was observed immediately after RXJ1615 at airmass 1.07, and with the same slit.
 
 \begin{figure}
 \includegraphics[scale=0.45]{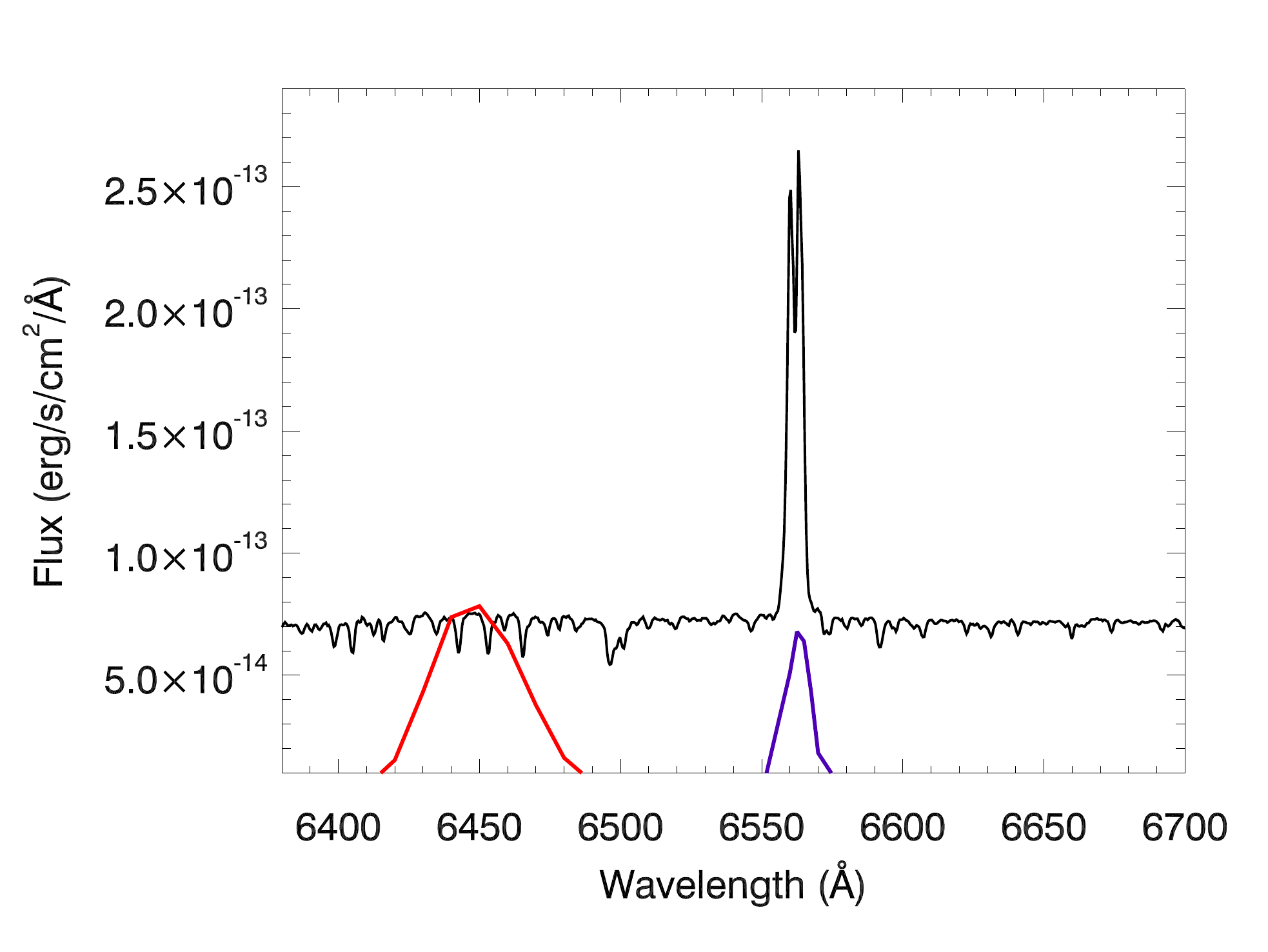}
 \caption{MagE spectrum of RXJ1615. We display the wavelength range containing the H$_{\alpha}$ emission line. The red and blue lines represent the transmission curves of the ZIMPOL Cnt\_Ha and N\_Ha filters, respectively.}\label{specMagE}
 \end{figure}
 
 The spectra were reduced with a dedicated pipeline. The processing includes bias subtraction and flat-field correction. We  used  thorium-argon (ThAr) exposures obtained after the science spectrum to wavelength-calibrate the data. We then flux-calibrated the science spectrum using the standard star. We used the spectrum to perform synthetic photometry, deriving an R-band magnitude of 11\,mag, close to the tabulated value of 11.2\,mag \citep[e.g.,][]{Makarov2007}.  

Figure~\ref{specMagE} shows part of the calibrated spectrum containing the H$_{\alpha}$ line, together with the ZIMPOL Cnt\_Ha and N\_Ha transmission curves. We measure a H$_{\alpha}$ line flux of 1.2$\pm$0.2 $\times$ 10$^{-12}$ erg/s/cm$^2$, and a continuum flux density of 7.2$\times$ 10$^{-14}$ erg/s/cm$^2$/\AA. The two values are consistent with the derived ZIMPOL data, validating the H$_{\alpha}$ line flux estimation procedure.

 {\bf LkCa~15}: This object was spectroscopically studied in detail by \citet{Whelan2015} using VLT/XSHOOTER data. The star was observed in six epochs displaying clear H$_{\alpha}$ variability. The H$_{\alpha}$ line flux measured in the different epochs varied between (0.7 -- 1.56)$\times 10^{-12}$ erg/s/cm$^2$. The derived H$_{\alpha}$ ZIMPOL flux lies within this interval, being very close to the strongest value.
 On the other hand, \citet{Manara2014} reported a factor $\sim$2 stronger H$_{\alpha}$ line flux of 3.1$\pm$0.1$\times 10^{-12}$ erg/s/cm$^2$.

  The estimated ZIMPOL accretion rate is 
  $ M_{\rm acc} \sim 2\times 10^{-9}$ $M_{\odot}$/yr. However, if we consider the minimum and maximum line fluxes measured in the works mentioned above, and the stellar parameters from Table~\ref{stars1}, we derive accretion rates of 
  $\sim 9\times 10^{-10} -  4.8\times 10^{-9}\,M_{\odot}$/yr.

{\bf TW~Hya}: The H$_{\alpha}$ line flux and profile of this target is very variable with rapid changes in timescales of days \citep[see e.g.,][]{Dupree2012}. The H$_{\alpha}$ line flux derived in the ZIMPOL observations (2.87$\pm$1.25$\times10^{-11}$ erg/s/cm$^{-2}$) is consistent within the uncertainties with the value reported by \citet{Manara2014}: 2.39$\pm$0.04$\times10^{-11}$ erg/s/cm$^{-2}$. Using the stellar parameters included in Table~\ref{stars1}, we estimate an accretion rate of $M_{\rm acc} \sim 5\times 10^{-9} M_{\odot}$/yr.


{\bf T Cha}: This object is known to be extremely variable both in the optical continuum and the H$_{\alpha}$ line \citep[see e.g.,][]{Alcala1993,Schisano2009,Cahill2019}. In fact, the line has been observed both in emission and in absorption. Previous works suggested that the variability could be related with strong circumstellar extinction due to the presence of dusty clumps in the inner disk. 

We estimate a very low H$_{\alpha}$ emission flux in the ZIMPOL data, resulting in an accretion rate of $M_{\rm acc} \sim 2 \times 10^{-10}\,M_{\rm \odot}$/yr. This value is  between the rate reported by \citet{Schisano2009}, estimated through the relationship between $M_{\rm acc}$ and the 10\% width of the H$_{\alpha}$ line (4$\times 10^{-9}\,M_{\rm \odot}$/yr), and the value estimated by \citet{Cahill2019} using the H$_{\alpha}$ line in emission observed in one of their analyzed spectra (8$\times 10^{-11}$ $M_{\odot}$/yr).

We note that we estimated a continuum flux density of $\sim$ 2$\times 10^{-13}$ erg/s/cm$^2$/\AA\  in the ZIMPOL data, which is comparable to the value derived by \citet{Cahill2019} for an extinction value of 1.2\,mag. Hence, we adopted this value for $A_v$.

{\bf HD163296:} The H$_{\alpha}$ emission of this object has been studied by different authors:
\citet{Wichitta2020} reported a H$_{\alpha}$ line flux of 1.35$\pm$0.05$\times$10$^{-10}$ erg/s/cm$^{-2}$, 
while \citet{Fairlamb2015} estimated a line flux of 1.06$\pm$0.09 $\times$ 10$^{-10}$ erg/s/cm$^{-2}$.
\citet{Mendigutia2013} studied data for HD163296  from five epochs  obtained on timescales from days to months.
We converted their provided maximum and minimum H$_{\alpha}$ luminosities into fluxes using their assumed distance of 130\,pc,  
obtaining line fluxes of 1.5 -- 2.0 $\times$ 10$^{-10}$ erg/s/cm$^{-2}$.

Using the ZIMPOL data, we report a H$_{\alpha}$ line flux of 0.82$\pm$0.36$\times$10$^{-10}$ erg/s/cm$^{-2}$, which is within the reported values in previous works within the uncertainties.
Using equations \ref{laccHAB} and \ref{maccHAB}, we estimate an accretion rate of 
9.8$\times$10$^{-8}$ M$_{\odot}$/yr. If we consider the minimum and maximum H$_{\alpha}$ line fluxes estimated for the object, and the updated stellar parameters included in Table~\ref{stars1}, the accretion rate can vary between 0.9 and 2.4 $\times$ 10$^{-7}$ M$_{\odot}$/yr.

\end{appendix}

\end{document}